\documentclass[journal,twoside,web]{ieeecolor}
\usepackage{generic}
\usepackage{amsmath,amssymb,amsfonts}
\usepackage{algorithmic}
\usepackage{graphicx}
\usepackage{algorithm,algorithmic}
\usepackage{textcomp}
\usepackage{booktabs}
\usepackage{subcaption}
\usepackage{siunitx}
\usepackage{float}
\usepackage{hyperref}
\newcommand{\mat}[1]{\ensuremath{{\mathbf{\MakeUppercase{#1}}}}}
\newcommand{\transpose}[1]{\ensuremath{{#1}^{\textsc{t}}}}
\newcommand{\trace}[1]{\ensuremath{\text{Tr}\left(#1\right)}}

\def\BibTeX{{\rm B\kern-.05em{\sc i\kern-.025em b}\kern-.08em
    T\kern-.1667em\lower.7ex\hbox{E}\kern-.125emX}}
\begin{document}
\title{EEG-based Decoding of Selective Visual Attention in Superimposed Videos}
\author{Yuanyuan Yao, Wout De Swaef, Simon Geirnaert, and Alexander Bertrand, \IEEEmembership{Senior Member, IEEE}
\thanks{This research is funded by the Research Foundation - Flanders (FWO) project No G081722N, junior postdoctoral fellowship fundamental research of the FWO (for S. Geirnaert, No. 1242524N), the European Research Council (ERC) under the European Union's Horizon 2020 research and innovation program (grant agreement No 101138304), Internal Funds KU Leuven (project IDN/23/006), and the Flemish Government (AI Research Program). Views and opinions expressed are however those of the author(s) only and do not necessarily reflect those of the European Union or the granting authorities. Neither the European Union nor the granting authorities can be held responsible for them.}
\thanks{All authors are with KU Leuven, Department
of Electrical Engineering (ESAT), STADIUS Center for Dynamical Systems,
Signal Processing and Data Analytics and with Leuven.AI - KU Leuven
institute for AI, Kasteelpark Arenberg 10, B-3001 Leuven, Belgium (e-mail: yuanyuan.yao@kuleuven.be, wout.deswaef@student.kuleuven.be, simon.geirnaert@kuleuven.be, alexander.bertrand@kuleuven.be).}
\thanks{S. Geirnaert is with KU Leuven, Department of Neurosciences,
Research Group ExpORL, Herestraat 49 box 721, B-3000 Leuven,
Belgium.}
}

\maketitle

\begin{abstract}
  Selective attention enables humans to efficiently process visual stimuli by enhancing important elements and filtering out irrelevant information. Locating visual attention is fundamental in neuroscience with potential applications in brain-computer interfaces. Conventional paradigms often use synthetic stimuli or static images, but visual stimuli in real life contain smooth and highly irregular dynamics. We show that these irregular dynamics can be decoded from electroencephalography (EEG) signals for selective visual attention decoding. To this end, we propose a free-viewing paradigm in which participants attend to one of two superimposed videos, each showing a center-aligned person performing a stage act. Superimposing ensures that the relative differences in the neural responses are not driven by differences in object locations. A stimulus-informed decoder is trained to extract EEG components correlated with the motion patterns of the attended object, and can detect the attended object in unseen data with significantly above-chance accuracy. This shows that the EEG responses to naturalistic motion are modulated by selective attention. Eye movements are also found to be correlated to the motion patterns in the attended video, despite the spatial overlap with the distractor. We further show that these eye movements do not dominantly drive the EEG-based decoding and that complementary information exists in EEG and gaze data. Moreover, our results indicate that EEG may also capture neural responses to unattended objects. To our knowledge, this study is the first to explore EEG-based selective visual attention decoding on natural videos, opening new possibilities for experiment design.
\end{abstract}


\section{Introduction} \label{sec:intro}
\label{sec:introduction}
\IEEEPARstart{I}{n} everyday life, humans are constantly exposed to a vast amount of visual information. To process this with limited resources, the brain has developed a mechanism known as selective visual attention, which enables individuals to prioritize stimuli of interest in the visual field while suppressing others \cite{carrasco2011visual}. Decoding selective visual attention has been a popular research topic in neuroscience and brain-computer interface communities, providing insights into the neural basis of attention and offering potential applications in various fields. For example, it can aid communication and control for individuals with severe paralysis \cite{kellyVSA2005, reichert2020impact}, diagnosis of attention and consciousness-related disorders \cite{liInformationbased2023, abiri2019decoding, montiVisualCognition2013}, rehabilitation of cerebral-visual impairment or cognitive deficits \cite{astrandSVA2014, debettencourt2015closed}, and optimization of streaming processes in virtual reality \cite{ozcinarVAA2019}.

\par Extensive research has been conducted on the mechanisms underlying selective visual attention. These studies have shown that, although the sensory representations of both attended and unattended stimuli are present in the visual field, the attended stimulus elicits stronger cortical responses \cite{moran1985selective, yantis2003cortical}. This modulation effect of attention on neural responses enables the neural-based decoding of selective visual attention. For example, Kelly et al. \cite{kellyVSA2005} successfully decoded covert left/right spatial attention from steady-state visual evoked potentials elicited by flickering stimuli. The classification was based on the amplitude of the evoked potentials at the flicker frequency, which was approximately doubled when the flickering stimulus was attended.

\par Apart from spatial locations, selective attention also frequently targets specific objects. Attention enhances the features of the attended object, such as its motion, color, or shape, even when attended and unattended objects are superimposed \cite{yantis2003cortical, tong2012decoding}. Early studies using functional magnetic resonance imaging (fMRI) have shown that it is possible to decode the category of the attended object in an image with superimposed objects from different categories \cite{debettencourt2015closed, niaziODO2014}. In more recent studies such as \cite{kellerAEC2022}, the categories of unattended objects were also decoded and compared with the decoding accuracy of attended objects, showing that the latter were more accurately decodable. These studies trained classifiers only on brain signals to decode the object category, whereas Horikawa et al. \cite{horikawaAMN2022} incorporated image features, decoding these features from the fMRI voxels using linear regression. Additionally, a deep generator network was appended after the regression model to generate images from the output features. The overall model predicted the features of the superimposed images and generated corresponding images that, as shown in the study, were similar to the attended object. Efforts have also been made to decode selective attention on superimposed images using electroencephalography (EEG) \cite{abiri2019decoding, grootswagersND2021}, as EEG has much more potential for real-world applications due to its affordability and portability. Additionally, the high temporal resolution of EEG enables the capture and analysis of neural responses not only to static images but also to images presented in rapid succession \cite{grootswagersND2021}.

\par Previous studies provide valuable insights into the neural basis of selective visual attention and demonstrate the feasibility of decoding selective attention based on brain signals. However, these studies primarily focus on synthetic stimuli and static or rapid serial images of various objects or scenes, which do not reflect the dynamic and continuous visual stimuli encountered in real life. This motivates us to explore natural and more dynamic visual stimuli: videos. EEG has a high temporal resolution such that it can capture the fast dynamics of neural responses elicited by the time-varying features within the video \cite{yao2024identifying}. Using EEG signals, we aim to decode the attended object in videos with two moving objects (persons) that spatially overlap. As we consider a naturalistic, free-viewing condition, the use of overlapping objects is crucial for eliminating the possibility that differences in neural responses are driven by different spatial locations of the two objects (relative to the focus of gaze) rather than by selective attention. This design enables us to confidently attribute the decoding results to attention-based modulation of neural signals. To the best of our knowledge, no study has yet attempted to decode selective visual attention based on EEG signals when viewing natural videos. This work may lay the foundation for more naturalistic experiment design in cognitive neuroscience and brain-computer interfaces, more patient-friendly assessments of attention-related disorders or cerebral-visual impairment, and attention tracking systems, e.g., for education and neuromarketing \cite{gavaret2023eeg, jamil2021cognitive}. It could also find applications in augmented reality, such as identifying whether the user's attention is focused on the (superimposed) virtual or real world \cite{si2018towards}.

\par In this work, the selective attention decoding is based on stimulus reconstruction, identifying the attended object by comparing the temporal correlations between the EEG responses and the features of the object(s) in the video. We choose this correlation-based paradigm since direct classification based on instantaneous features could overfit to confounds such as trial-dependent EEG feature shifts \cite{li2020perils, rotaru2024we}. This paradigm has been widely applied in neural tracking to speech and auditory attention decoding \cite{biesmans_auditory-inspired_2017, geirnaert_electroencephalography-based_2021, puffayRelatingEEGContinuous2023}, where a common practice is to correlate the EEG signals with, for example, the envelope of the speech signals and decode the attended speaker as the one with the highest correlation. We hypothesize that a similar differentiated neural processing for attended and unattended stimuli exists in the visual sensory system using an analogous feature in the visual domain. One that, like the auditory envelope, correlates with EEG signals and is selectively enhanced through attentional modulation. One candidate of such a feature is the motion-encoding object-based optical flow proposed in \cite{yao2024identifying}, which extracts a time series that contains the average optical flow within the object at each time point, and was found to have significant correlations with EEG signals. However, in \cite{yao2024identifying}, the experiments used single-object videos without manipulating attention,  leaving it unclear how attention and competing stimuli influence the results. In this study, we show that the correlation between the EEG responses and this feature is indeed modulated by attention, allowing us to decode selective visual attention. Since participants are allowed to freely watch the videos without fixating on a specific point (though tracking only one of two center-aligned superimposed video objects), we also investigate the possibility of decoding selective attention by correlating the per-object optical flow time series with the eye movements, comparing their performance with EEG-based decoding.

\par The rest of this paper is organized as follows: Section \ref{sec:materials_methods} details the experimental setup, data preprocessing and feature extraction, introduces the analysis tools, and describes the evaluation tasks and the practicalities of our implementation. Section \ref{sec:results} presents the results and their implications, with a more in-depth discussion in Section \ref{sec:discussion}. Section \ref{sec:conclusion} concludes the paper.

\section{Materials and Methods} \label{sec:materials_methods}

\subsection{Stimuli} \label{sec:stimuli}

\par This study primarily focuses on the neural decoding of selective visual attention when viewing natural videos with two moving objects in a naturalistic, free-viewing condition. To avoid confounds of audio in neural processing, the videos are muted when presented to the participants. There is an important restriction on the experiment videos: they must be single-shot, i.e., with static camera angles and no scene changes. It is motivated by previous studies that have shown that the discontinuities due to shot cuts in videos can elicit strong neural responses \cite{yao2024identifying, herbec_differences_2015, nentwich2023semantic}, which are absent in natural visual stimuli, and which may lead to over-optimistic decoding performance \cite{yao2024identifying}.

\par
\begin{figure*}
  \centering
  \includegraphics[width=.9\linewidth]{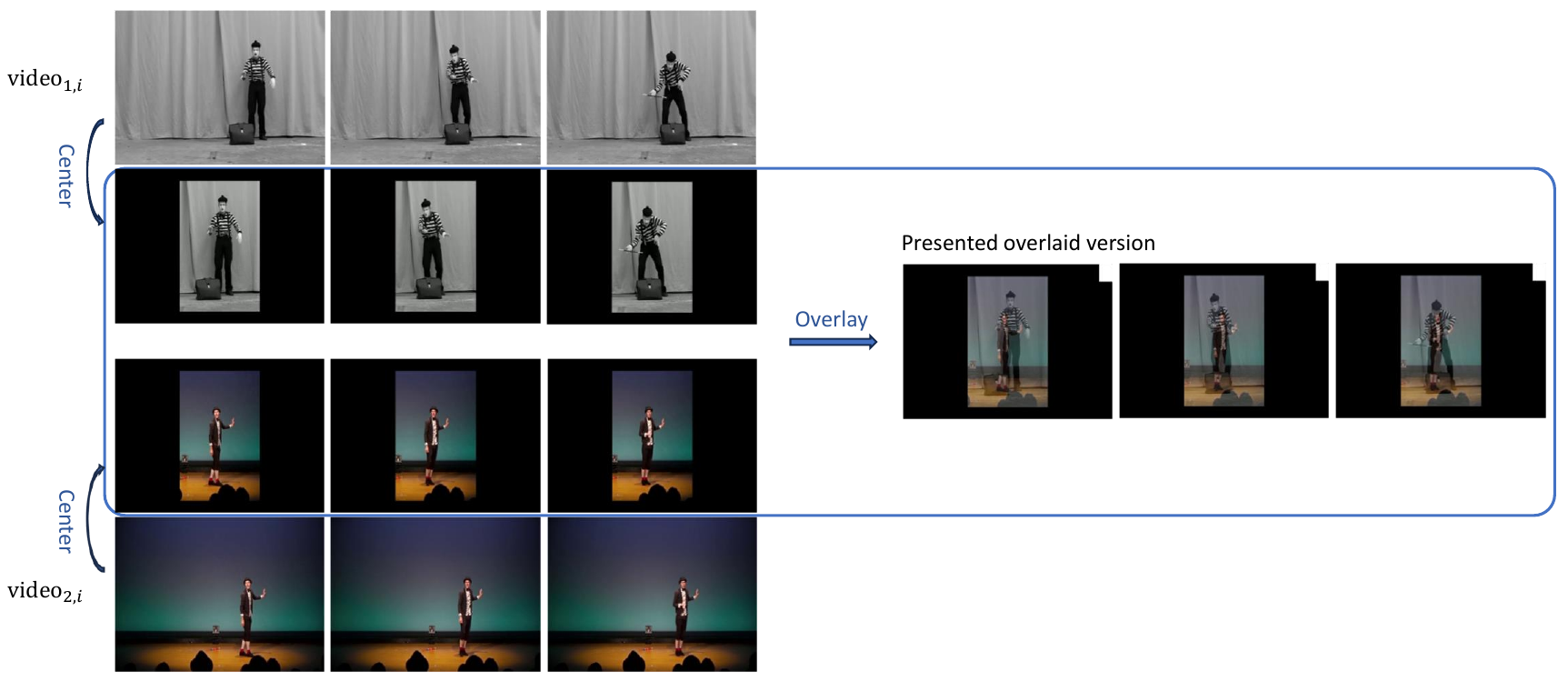}
  \caption{An illustration of creating superimposed videos. The objects in two single-shot videos are centered and overlaid with 50\% transparency. A white box is inserted in the top right corner to indicate the content-playing stage. For a detailed timeline of the experimental procedure, please refer to Fig. \ref{fig:exp_video_creation}.}
  \label{fig:overlay_videos}
\end{figure*}

 We create video stimuli by superimposing pairs of single-shot videos, each containing a single moving person. We superimpose the videos rather than placing them side by side to ensure that any modulation of the correlation between neural responses and object motion is not confounded by the location of the objects. In a pilot experiment, we found that spatially separating two objects led participants to shift their gaze to the attended object, leaving the unattended object in peripheral vision. This resulted in much weaker neural responses for the unattended object, making the selective attention decoding problem rather trivial. Inspired by studies using superimposed images \cite{debettencourt2015closed, niaziODO2014, kellerAEC2022, horikawaAMN2022, grootswagersND2021}, we center-align the objects in both videos and superimpose them with $50\%$ transparency (Fig. \ref{fig:overlay_videos}). This design ensures that both objects remain simultaneously visible while occupying the same spatial location, creating a more challenging paradigm that represents a worst-case scenario for selective attention decoding. Additionally, this approach prevents simply decoding the attended object based on gaze point coordinates, reducing potential confounds from eye movements.

\par $14$ single-shot, single-object videos with an average length of \SI{305}{\second} are selected, most of which are inherited from a previous study \cite{yao2024identifying}. The frame rate is \SI{30}{\hertz}, and the resolution is $1920\times1080$. The content of these videos includes a person performing a specific stage act, such as dancing, acrobatics, magic shows, and mime shows. The $14$ videos are paired into $7$ pairs: $(\text{video}_{1,i}, \text{video}_{2,i}), i \in \{1, 2,...,7\}$. The videos in each pair are not necessarily from different content categories but are distinct enough both visually and in terms of motion patterns, such that it is relatively easy for a participant to focus on one object while ignoring the other. 

\par
\begin{figure*}[htbp]
  \centering
  \includegraphics[width=\linewidth]{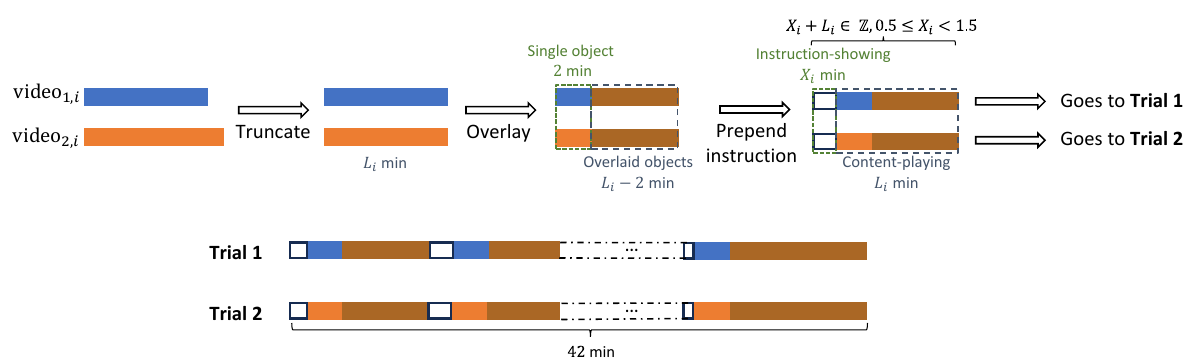}
  \caption{  An illustration of creating experimental videos: The original video pairs are truncated, overlaid from the second minute, and prepended with instruction frames. The videos in each pair are assigned to different trials.}
  \label{fig:exp_video_creation}
\end{figure*}

 Fig. \ref{fig:exp_video_creation} illustrates the procedure of creating the experiment stimuli from these $7$ pairs. In each pair, we truncate the videos to the minimum length of the two, and superimpose them with $50\%$ transparency, except for the first two minutes. In these first two minutes, only the video of the attended object is visible (i.e., with a $100\%$ transparency for the unattended video). The transition from single video to $50\%$-$50\%$ superimposed videos is made smooth by linearly changing the transparency over two seconds. Each video pair is presented twice, where the attended object switches in both presentations. This means that the first two minutes (showing only the attended object) is different in both presentations, yet the remaining part (showing $50\%$-$50\%$ superimposed videos) is exactly the same stimulus in both presentations.

\par Instruction frames are prepended in each video. These instruction frames contain a QR code for synchronization and an instruction text asking participants to always focus on the first object presented during the first two minutes of the video. The number of instruction frames is set to ensure they last longer than $30$ seconds and make the total video length a multiple of one minute. A progress bar is embedded to indicate the start of the video playback. The experiment consists of two trials of $42$ minutes during which each video pair is presented once (randomized across participants). In the second trial, the $7$ pairs are presented again, but with attention to the other object.

\subsection{Participants and data acquisition} \label{sec:participants_data_acquisition}

\par $19$ young, healthy adults participated in the experiment. All have normal or corrected-to-normal vision and no history of neurological disorders. The study is approved by the KU Leuven Social and Societal Ethics Committee, and all participants provided written informed consent before the experiment.


\par During the experiment, participants are seated comfortably in a quiet, separate room, approximately \SI{90}{\centi\meter} from the screen. The stimuli (videos) are presented on a $22$-inch monitor with a resolution of $1920\times1080$ pixels and a refresh rate of \SI{60}{\hertz}. Participants are instructed to watch the videos attentively and naturally, and to minimize unnecessary movements for better data quality. While participants watch the videos, their EEG data are recorded using a BioSemi ActiveTwo system (BioSemi B.V., Amsterdam) with $64$ channels and a sampling rate of \SI{2048}{\hertz}. Eye movements are coregistered with EEG using four electrooculogram (EOG) electrodes placed above and below the right eye and on the outer canthi of both eyes. Participants also wear a NEON eye tracker (Pupil Labs GmbH, Berlin) to acquire gaze data at \SI{200}{\hertz}. Four markers are placed around the screen for defining the surface that the gaze data are mapped to.

\par The experiment videos consist of interleaving instruction-showing and content-playing stages, as detailed in Section \ref{sec:stimuli}. A small box is embedded in the top right corner of the videos without occluding any content, serving as an indicator of the two different stages. The box is black during the instruction stage and turns white when the content starts to play, which is captured by a photodiode fixed at the corresponding region and connected to the EEG recorder. Synchronization between the EEG data and the video stimuli is achieved by detecting the upper edges of the photodiode signal. The embedded box is covered with black tape to prevent distraction.

\par To synchronize the eye tracker data with the video stimuli, a QR code is encoded during the instruction stage and is detected post hoc from the videos recorded by the world camera of the eye tracker. The time points at which the QR code appears are identified and aligned with the corresponding time points during the instruction stage. The first time point when the QR code disappears from the video is the synchronization point between the eye tracker data and the video stimuli.

\par The three data sources (EEG recorder, eye tracker, and video stimuli) are thus synchronized. This synchronization is performed per short video clip ($4\sim 8$ minutes) in each trial, rather than only once at the beginning, to prevent non-negligible time lags caused by differences in the time clocks of each device.


\subsection{Data preprocessing} \label{sec:data_preprocessing}

\par The EEG and EOG data are first segmented based on the photodiode signals, which indicate the start of each content-playing stage. Basic preprocessing is applied to each segment, including interpolation of bad channels, average re-referencing, high-pass filtering with a cutoff frequency of \SI{0.5}{\hertz} to remove drifts, notch filtering at \SI{50}{\hertz} to remove powerline noise, and downsampling to \SI{30}{\hertz} (including anti-aliasing) to match the video frame rate. The filters are zero-phase and thus no delays are introduced.

\par From the eye tracker, we export the gaze coordinates mapped to the screen surface, the start and end points of fixations, and the time points of eye blinks. Saccades are identified as the end of fixations and are represented as a binary time series, indicating whether a saccade occurs at each frame. Eye blinks in the gaze data are linearly interpolated, and the gaze data are downsampled to \SI{30}{\hertz}.

\par Eye movements can also be informative, as participants' gaze may track the movement pattern of the attended object. We therefore extract a feature from both the gaze and the EOG data that is representative for the velocity of the eye movements:
\begin{equation}
  \text{velocity} = \sqrt{(a(t)- a(t-1))^2+(b(t)- b(t-1))^2},
  \label{eq:velocity}
\end{equation}
where $a(t)$ and $b(t)$ represent the horizontal and vertical gaze coordinates or the horizontal and vertical EOG channels at time $t$, respectively.

\par Overall, six data modalities are extracted for further analysis, as summarized in Table \ref{tab:data_mod}. The data are further divided into two sets based on whether a single-object video or an superimposed-object video is playing. The fade-in periods are excluded from the analysis. Additionally, the first and last second of each video segment are also excluded to avoid potential effects caused by video onset and offset. In the end, the single-object dataset contains $19 \text{ subjects } \times 14 \text{ videos } \times 2 \text{ minutes}$ of data. The superimposed-object dataset contains $19$ subjects, $14$ videos with an average video length of \SI{166}{\second} and a standard deviation (STD) of \SI{76}{\second}, totaling approximately $19 \text{ subjects } \times 38 \text{ minutes}$ of data. The instruction-showing stage at the beginning of each video also serves as a short break for the participants and is a total of $18$ minutes long.

\begin{table}[ht]
  \centering
  \caption{  Data modalities extracted from the recorded data.}
  \renewcommand{\arraystretch}{1.5}
  \begin{tabular}{r|l}
  \textbf{Modality} & \textbf{Abbreviation} \\
  \hline
  64-channel EEG signal  & $EEG$  \\
  4-channel EOG signal  & $EOG$  \\
  2D gaze coordinate  & $GAZE$ \\
  Saccade (binary time series)  &   $SACC$  \\
  Velocity \eqref{eq:velocity} calculated from gaze &  $GAZE\_V$  \\
  Velocity \eqref{eq:velocity} calculated from EOG &  $EOG\_V$  \\
  \end{tabular}
  \label{tab:data_mod}
  \end{table}

\subsection{Video feature extraction} \label{sec:video_feature_extraction}

\par Our approach for decoding selective attention is by identifying the temporal correlation between the dynamics in the video and the stimulus-following neural responses that follow these time-varying features. However, video data are high-dimensional, leading to an explosion of model parameters if it would be used in its raw format. Therefore, it is crucial to first extract relevant features that elicit strong neural responses in order to reduce data dimensionality. In \cite{yao2024identifying}, object-based optical flow (\textit{ObjFlow}) and object-based temporal contrast (\textit{ObjTempCtr}) were found to be correlated with EEG signals. We select \textit{ObjFlow} for this study, as the performance of both features is comparable.

\par \textit{ObjFlow} is defined as the average optical flow magnitude within the object of interest:
\begin{equation}
\text{ObjFlow} = \frac{1}{|\mathcal O|}\sum_{\mathbf z \in \mathcal O} \left | \mathbf v(\mathbf z, t) \right |,
\label{eq:OF_def}
\end{equation}
where $\left | \mathbf v(\mathbf z, t) \right |$ denotes the magnitude of pixel velocity at position $\mathbf z$, time $t$, and $|\mathcal O|$ represents the number of pixels in the object mask. In practice, the object mask is obtained by applying a pre-trained object segmentation model, Mask R-CNN \cite{he_mask_2020}. The optical flow is calculated using the Gunnar-Farneback method \cite{goos_two-frame_2003}. Features are extracted from each video before superimposing them, ensuring that potential confounds from artificial overlapping effects are avoided. All videos are downsampled to $640\times360$ pixels to reduce computational cost before feature extraction.

\subsection{Correlation analysis} \label{sec:correlation_analysis}

\par Correlation analysis can be conducted on two or more views to measure their temporal coupling, optionally controlling for the effects of certain variables. In this section, we briefly review canonical correlation analysis and its two extensions, partial canonical correlation analysis and generalized canonical correlation analysis, and explain their application to our data. All data modalities and extracted features are centered before correlation analysis.

\subsubsection{Canonical Correlation Analysis (CCA)} \label{sec:cca}

\par CCA is a method for finding correlations between two sets of variables \cite{hotelling1992relations}. In this study, it is used to quantify the correlations between the video stimuli and the various data modalities introduced in Section \ref{sec:data_preprocessing} and Table \ref{tab:data_mod}. When correlating a data modality (e.g., EEG signals) $\mathbf x(t) \in \mathbb R^{D_x}$ with video features $\mathbf y(t) \in \mathbb R^{D_y}$, CCA finds linear maps $\mathbf w_x\in \mathbb R^{D_x}$ and $\mathbf w_y\in \mathbb R^{D_y}$ that maximize the correlation between the transformed signals $\transpose{\mathbf w}_x \mathbf x(t)$ and $\transpose{\mathbf w}_y \mathbf y(t)$. Notably, this process inherently filters out EEG artefacts that are not systematically correlated with video features. Mathematically, this can be expressed as the following optimization problem \cite{hotelling1992relations}:
\begin{equation}
  \begin{aligned}
  & \underset{\mathbf{w}_x, \mathbf{w}_y}{\text{maximize}}
  & & \mathbb E\{[\transpose{\mathbf w}_x \mathbf x(t)][\transpose{\mathbf w}_y \mathbf y(t)]\}\\
  & \text{subject to}
  & & \mathbb E\{[\transpose{\mathbf w}_x \mathbf x(t)]^2\}=1, \\
  &&& \mathbb E\{[\transpose{\mathbf w}_y\mathbf y(t)]^2\} = 1,
  \end{aligned}
  \label{eq:cca}
  \end{equation}
where $\mathbb E\{\cdot\}$ denotes the expectation operator. Correlations between neighboring samples can also be incorporated by extending $\mathbf x(t)$ and $\mathbf y(t)$ with $L_x-1$ and $L_y-1$ time-lagged copies, respectively, which also allows to automatically correct for relative time delays between $\mathbf x(t)$ and $\mathbf y(t)$. $\mathbf w_x$ and $\mathbf w_y$ then become spatial-temporal, with dimensions $D_x L_x$ and $D_yL_y$. Solving problem \eqref{eq:cca} requires estimating the covariance matrices $\mathbf R_{xy}=\mathbb E\{\mathbf x(t)\transpose{\mathbf y (t)} \}\in \mathbb R^{D_xL_x\times D_y L_y}$, $\mathbf R_{xx}=\mathbb E\{\mathbf x(t)\transpose{\mathbf x (t)} \}\in \mathbb R^{D_xL_x\times D_xL_x}$, and $\mathbf R_{yy}=\mathbb E\{\mathbf y(t)\transpose{\mathbf y (t)} \}\in \mathbb R^{D_y L_y\times D_y L_y}$, which can be approximated by the sample covariance matrices. 

\par While \eqref{eq:cca} only aims to find a single canonical component $\mathbf w_x$ and $\mathbf w_y$, often higher-order canonical components are jointly estimated. Let $\mathbf{w}_x^k$ and $\mathbf{w}_y^k$ denote the $k$-th order canonical components, and $\transpose{\mathbf{x}(t)} \mathbf{w}_x^k$ and $\transpose{\mathbf{y}(t)} \mathbf{w}_y^k$ represent the $k$-th canonical directions. These components, $\mathbf{w}_x^k$ and $\mathbf{w}_y^k$, are linear maps applied to the data such that the transformed signals are orthogonal to all preceding canonical directions and are maximally correlated. In compact form, the multi-component version of \eqref{eq:cca} can be formulated as:
\begin{equation}
  \begin{aligned}
  & \underset{\mathbf{W}_x, \mathbf{W}_y}{\text{maximize}}
  & & \trace{\transpose{\mat{W}}_x\mat{R}_{xy}\mat{W}_y} \\
  & \text{subject to}
  & & \transpose{\mat{W}}_x\mat{R}_{xx}\mat{W}_x = \mat{I}_K, \\
  &&& \transpose{\mat{W}}_y\mat{R}_{yy}\mat{W}_y = \mat{I}_K,
  \end{aligned}
  \label{eq:cca_multi}
\end{equation}
where $K$ is the number of components, $\mathbf I_K$ is the identity matrix of size $K$, $\mathbf{W}_x \in \mathbb R^{D_xL_x\times K}$ and $\mathbf{W}_y\in \mathbb R^{D_yL_y\times K}$ store the canonical components as columns, and $\trace{\cdot}$ denotes the trace of a matrix. It can be shown that the solution to \eqref{eq:cca_multi} can be obtained by solving the following generalized eigenvalue decomposition (GEVD) problem \cite{corrochano2005eigenproblems}:
\begin{equation}
  \begin{bmatrix}
    \mathbf R_{xx} & \mathbf R_{xy} \\
    \mathbf R_{yx} & \mathbf R_{yy}
    \end{bmatrix} \begin{bmatrix}
      \mathbf W_{x} \\
      \mathbf W_{y} 
      \end{bmatrix} = \begin{bmatrix}
        \mathbf  R_{xx} & \mathbf 0 \\
        \mathbf 0 & \mathbf  R_{yy}
        \end{bmatrix}\begin{bmatrix}
        \mathbf W_{x} \\
        \mathbf W_{y} 
        \end{bmatrix}\mathbf \Lambda, \label{eq:cca_gevd}
\end{equation}
where $\mathbf \Lambda \in \mathbb R^{K\times K}$ is a diagonal matrix containing the generalized eigenvalues (GEVLs). The first $K$ canonical components are the generalized eigenvectors (GEVCs) corresponding to the $K$ largest GEVLs. The components (columns of $\mathbf W_x$ and $\mathbf W_y$) are rescaled to satisfy the constraints in \eqref{eq:cca_multi}.

\subsubsection{Partial Canonical Correlation Analysis (PCCA)} \label{sec:pcca}

\par PCCA was proposed by Rao in \cite{rao1969partial} as an extension of CCA to account for the effects of confounding variables $\mathbf{c} \in \mathbb{R}^{D_cL_c}$, where $D_c$ represents the dimension of the confounds and $L_c$ denotes the number of time-lagged copies. Consider the problem discussed in Section \ref{sec:cca}, where we aim to quantify the correlation between EEG signals and video features. Eye movements, often considered artefacts in EEG signals, may also correlate with video features, as specific patterns in the video may provoke particular eye movements. Therefore, it may be necessary to control for the effects of eye movements, which are captured by EOG signals.

\par PCCA involves one additional step compared to CCA: removing the effects of the confounds $\mathbf c$ from $\mathbf x$ and $\mathbf y$ by linear regression. Aggregating the samples of $\mathbf x$, $\mathbf y$, and $\mathbf c$ into matrices $\mathbf X \in \mathbb R^{T\times D_xL_x}$, $\mathbf Y \in \mathbb R^{T\times D_yL_y}$, and $\mathbf C \in \mathbb R^{T\times D_cL_c}$, respectively, the residuals can be written as:
\begin{subequations}
  \begin{align}
    \mathbf X_r &= \mathbf X - \mathbf P_c \mathbf X, \\
    \mathbf Y_r &= \mathbf Y - \mathbf P_c \mathbf Y,
  \end{align}
  \label{eq:residuals}
\end{subequations}
where $\mathbf P_c = \mathbf C(\transpose{\mathbf C}\mathbf C)^{-1}\transpose{\mathbf C}$ is the projection matrix onto the column space of $\mathbf C$. The residuals $\mathbf X_r$ and $\mathbf Y_r$ are then fed to the input of the CCA method. 

\subsubsection{Generalized Canonical Correlation Analysis (GCCA)} \label{sec:gcca}

\par GCCA is a generalization of CCA that can handle more than two views, making it useful for applications such as finding coherent EEG signals across multiple subjects. Two well-known formulations of GCCA are SUMCORR and MAXVAR \cite{kettenring1971canonical}. In this study, we select MAXVAR because the SUMCORR formulation does not have a closed-form solution \cite{fu2016efficient}.

\par MAXVAR-GCCA optimizes the decoders $\{\mathbf{W}_n\}_{n=1}^N$ applied to different views $\{\mathbf{X}_n\}_{n=1}^N$ to minimize the pairwise distances between the transformed views. An auxiliary variable $\mathbf S$ is introduced to represent the shared subspace among the views, and the optimization problem is formulated as:
\begin{equation}
  \begin{aligned}
  & \underset{\mat{W}_1,...,\mat{W}_N,\mat{S}}{\text{minimize}}
  & & \sum_{n=1}^N\left\|\mathbf{S}-\mathbf{X}_n \mathbf{W}_n\right\|_{\mathrm F}^2 \\
  & \text{subject to}
  & & \transpose{\mathbf{S}} \mathbf{S}=\mathbf{I}_K.
  \end{aligned}
  \label{eq:gcca}
\end{equation}
For the joint analysis of EEG signals from all subjects, the views $\mathbf{X}_n \in \mathbb{R}^{T \times D_x L_x}$ represent the EEG signals from different subjects, the matrices $\mathbf{W}_n \in \mathbb{R}^{D_x L_x \times K}$ are the per-subject decoders applied to these signals, and the shared subspace $\mathbf S\in \mathbb R^{T\times K}$ can be interpreted as the coherent EEG components across subjects. Other modalities in Table \ref{tab:data_mod} can be analyzed in a similar way.

\par Denote the covariance matrix between two views $\mathbf X_i, \mathbf X_j$ as $\mathbf R_{ij}$. It can be shown that the solution to \eqref{eq:gcca} can again be written as a GEVD problem similar to \eqref{eq:cca_gevd} \cite{geirnaert2023stimulusInformed}:

\begin{equation}
  \mathbf R \mathbf W = \mathbf D \mathbf W \mathbf \Lambda,
\end{equation}
where
\begin{align}
    \mathbf W &= \begin{bmatrix}
      \mathbf W_{1} \\
      \vdots\\
      \mathbf W_{N} 
      \end{bmatrix}, \quad 
    \mathbf R =  \begin{bmatrix}
      \mathbf R_{11} & \cdots & \mathbf R_{1N} \\
      \vdots & \ddots & \vdots \\
      \mathbf R_{N1} & \cdots & \mathbf R_{NN}
      \end{bmatrix}, \notag \\
    \mathbf D &= \begin{bmatrix}
      \mathbf  R_{11} & \mathbf 0 & \cdots & \mathbf 0 \\
      \mathbf 0 & \mathbf R_{22} & \cdots & \mathbf 0 \\
      \vdots & \vdots & \ddots & \vdots \\
      \mathbf 0 & \mathbf 0 & \cdots & \mathbf R_{NN}
      \end{bmatrix}.
  \end{align}
The columns of $\mathbf W$ are the GEVCs corresponding to the $K$ largest GEVLs. The shared subspace $\mathbf S$ can be obtained as the sum of the transformed views:
\begin{equation}
  \label{eq:gcca_shared}
  \mathbf S = \sum_{n=1}^N \mathbf X_n \mathbf W_n,
\end{equation}
with scalings applied to each column of $\mathbf{S}$ to ensure that $\mathbf{S}^\top \mathbf{S} = \mathbf{I}_K$.

\par Analogous to PCCA, the effects of confounds such as eye movements can be removed from each view by regressing out the confounds and then applying GCCA to the residuals. When performing group-level analysis using GCCA, inter-subject correlation (ISC) can be used to assess the overall similarity of the extracted components across different views \cite{dmochowski2012correlated}. ISC is defined as the average pairwise correlation between all component pairs:
\begin{equation}
  \text{ISC}_k = \frac{2}{N(N-1)}\sum_{i=1}^{N-1}\sum_{j=i+1}^N \text{corr}(\mathbf X_i \mathbf w_i^k, \mathbf X_j \mathbf w_j^k),
\label{eq:isc}
\end{equation}
where $\text{corr}(\cdot, \cdot)$ denotes the Pearson correlation, and $\mathbf w_i^k$ and $\mathbf w_j^k$ are the $k$-th columns of $\mathbf W_i$ and $\mathbf W_j$, respectively.


\subsection{Evaluation} \label{sec:evaluation_tasks}

\par The correlations obtained from (P)CCA provide insight into how well the data modalities and video features are temporally coupled. On the other hand, the ISCs calculated from GCCA indicate how well neural responses or eye movements are synchronized across subjects. In addition to these measures, we also perform two tasks to evaluate the decodability of attended objects, which is the primary focus of this study: a selective visual attention decoding (SVAD) task and a match-mismatch (MM) task.

\par The objective of both decoding tasks is to distinguish the attended video segment from an imposter (unattended or mismatched segment) using various data modalities. The only difference lies in whether the imposter is the observed but unattended competing video segment (SVAD) or an unobserved non-competing segment from a different time point in the same test set (MM). Since these tasks are highly similar but with different inputs, they can be tackled using the same decoding method. For example, with EEG data, the EEG decoders $\mathbf{W}_x$ and stimulus encoders $\mathbf{W}_y$ are trained on the EEG data and attended/matched video features using (P)CCA. During the testing phase, the previously obtained encoders and decoders are applied to the held-out test data, and the target video segment is identified by selecting the video that shows the strongest correlation with the EEG across the CCA components. The chance level for both tasks is $50\%$.


\par Intuitively, SVAD is more challenging because information from the unattended stimuli might also be decodable using the trained filters, making discrimination between the attended and unattended stimuli more difficult. In contrast, MM is easier since the imposter is not observed by the participant and can therefore not generate any correlated signal components in any of the recorded modalities. Therefore, evaluating both tasks together not only indicates how well the attended stimuli can be decoded but also provides insights into whether the analyzed data modality captures information from the unattended stimuli.

\subsection{Practicalities} \label{sec:practicalities}

\paragraph{Cross-validation} The accuracies and correlations reported in the following sections are cross-validated using a leave-one-pair-out scheme. Specifically, data from one video pair are left out for testing (this includes both presentations of the same pair), while data from the remaining pairs are used for training. This process is repeated $7$ times, corresponding to the $7$ video pairs, and the results are averaged across all pairs. In the single-object dataset, the training set and test set have a fixed length of \SI{24}{\minute} and \SI{4}{\minute}, respectively ($2 \times 2$ \SI{}{\minute} per pair). In the superimposed-object dataset, the training set has an average length of \SI{33.2}{\minute} (STD = \SI{2.5}{\minute}), and the test set has an average length of \SI{5.5}{\minute} (STD = \SI{2.5}{\minute}).

\paragraph{Time lags} The number of time lags in the CCA procedures are set in accordance with \cite{yao2024identifying}. Specifically, the video feature encoders have $L_y = 15$ lags, capturing video features from approximately the past \SI{0.5}{\second} to the current time point. For EEG decoders, the (P)CCA model uses $L_x = 3$ lags centered around the current time point, spanning approximately from \SI{-33}{\milli\second} to \SI{33}{\milli\second}, while the GCCA model employs $L_x = 5$ lags, covering approximately \SI{-67}{\milli\second} to \SI{67}{\milli\second}. These numbers are also applied to other data modalities in Table \ref{tab:data_mod}, as a grid search indicates that the results are not highly sensitive to the choice of the number of time lags.

\paragraph{Classifier} Multiple components can be extracted from (P)CCA, with each canonical component pair having a corresponding correlation value. To measure the ``closeness" of the data to the video candidates, we sum the correlations of the first two component pairs, as these often show statistically significant correlation values \cite{yao2024identifying}. The video segment with the higher score is then identified as the attended video. Although simple classifiers such as support vector machines or random forests can be applied to the complete set of obtained correlations, they do not yield significant improvements, and therefore the added complexity and increased risk for overfitting is not justified.

\paragraph{Statistical tests} We assess the significance of the correlations using a permutation test with phase scrambling. In phase scrambling, the phase components in the frequency domain are randomized to disrupt the temporal structure of the data while preserving the power spectrum \cite{prichard1994generating}. A null distribution of correlations is generated by repeating the correlation analysis on the phase-scrambled data $500$ times per fold. P-values are calculated as the proportion of correlations in the null distribution that are more extreme than the observed correlation (two-tailed). Note that this provides a relatively rigorous bound since the null distribution is computed from data with the same power spectrum. For decoding tasks, we assess whether the decoding accuracy is significantly above chance using a similar permutation approach. Test EEG trials are circularly shifted by a random number of trials to break their temporal alignment with the motion features. The null distribution is constructed by repeating the decoding process 100 times per subject using such shifted data, yielding a total of 1900 accuracy values. P-values are calculated using the same method as in the significance test for correlations. A correlation or accuracy is considered significant if it exceeds the threshold (significance level) corresponding to a p-value of $0.05$, which is the $97.5$th percentile of the null distribution. For comparing performance between different tasks or data modalities, we employ the Wilcoxon signed-rank test. When multiple comparisons are involved, p-values are adjusted using the Benjamini-Hochberg (BH) method \cite{benjamini2001control}. Performance differences are considered significant if the (adjusted) p-value is less than $0.05$.

\section{Results} \label{sec:results}

\subsection{Correlations are modulated by attention} \label{sec:correlation_modulation}

\par
\begin{figure*}
  \centering
  \includegraphics[width=.8\linewidth]{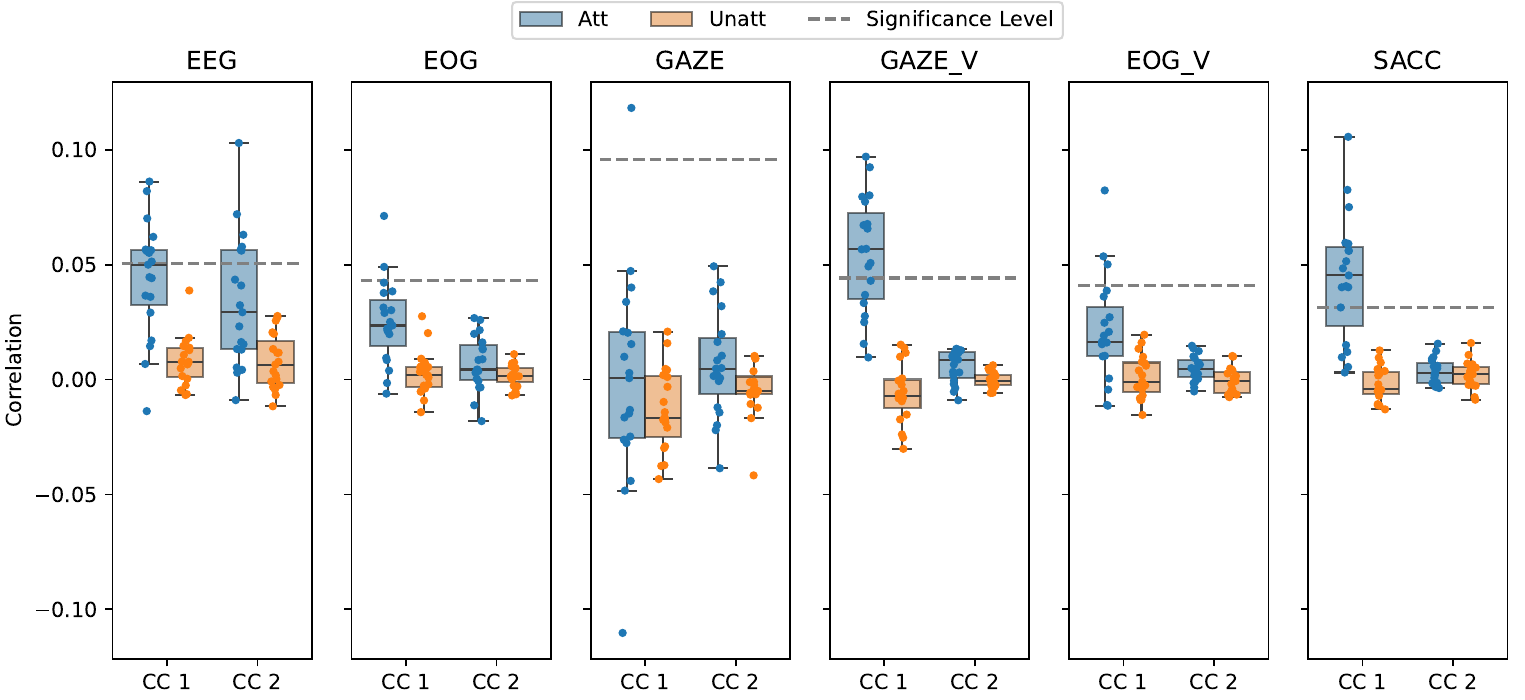}
  \caption{  The correlations of each data modality with the \textit{ObjFlow} features of the attended and unattended object in the superimposed-object dataset are shown in blue and orange, respectively. ``CC 1" and ``CC 2" denote the first and second canonical components. The dots represent the mean correlations of individual subjects across folds. The boxes indicate the median and the interquartile range, while the whiskers extend to the most extreme data points not considered outliers. The dashed lines represent the significance level pooled across subjects and components.}
  \label{fig:correlation_modulation}
\end{figure*}

 A core assumption in our method is that the correlations between the video features and the collected data modalities are modulated by attention. If this assumption holds, the attended object can be identified by comparing the relative strengths of these correlations between the attended and unattended objects. In this experiment (using the superimposed-object dataset), the encoders and decoders are trained on the \textit{ObjFlow} feature of the attended object and each data modality using CCA. The correlations are then computed on the test set for the \textit{ObjFlow} features of both the attended and unattended objects. The obtained correlation coefficients for the first two canonical components are shown in Fig. \ref{fig:correlation_modulation}.

\par The first observation is that the correlations with attended features (in blue) are generally higher than those with unattended features (in orange), especially for the first canonical component. Therefore, we can conclude that correlations are modulated by attention, justifying the design of our classifier (Section \ref{sec:practicalities}). Moreover, the correlations with unattended features are non-significant for all modalities, whereas for the attended case, the significance level is around or below the median for the modalities $EEG$, $GAZE\_V$, and $SACC$. It is also worth noting that the correlations between different modalities are not directly comparable, as the significance levels differ due to the different spectral characteristics of each modality. Their performance can be better compared in specific tasks, as specified below.

\subsection{Selective visual attention is decodable from EEG and eye movements} \label{sec:VAD_res}

\par Since the assumption that correlations are modulated by attention holds, identifying the attended video segment based on these correlations appears to be feasible. To evaluate how well the attended video segment can be decoded from the data, we perform the SVAD task on the superimposed-object dataset as described in Section \ref{sec:evaluation_tasks} and estimate the decoding accuracy using bootstrapping. Specifically, in each cross-validation fold, 30-second test segments are randomly sampled $V_l/3$ times, where $V_l$ is the length of the test set in seconds. The number of test segments is approximately $110$ per fold on average. Over each \SI{30}{\second} test segment, we compute the correlation between the tested modality and the \textit{ObjFlow} feature of both the attended and unattended object, and select the one exhibiting the highest correlation. The decoding accuracy is calculated as the proportion of times the attended video segment is correctly identified.

\par
\begin{figure}
  \centering
  \includegraphics[width=\linewidth]{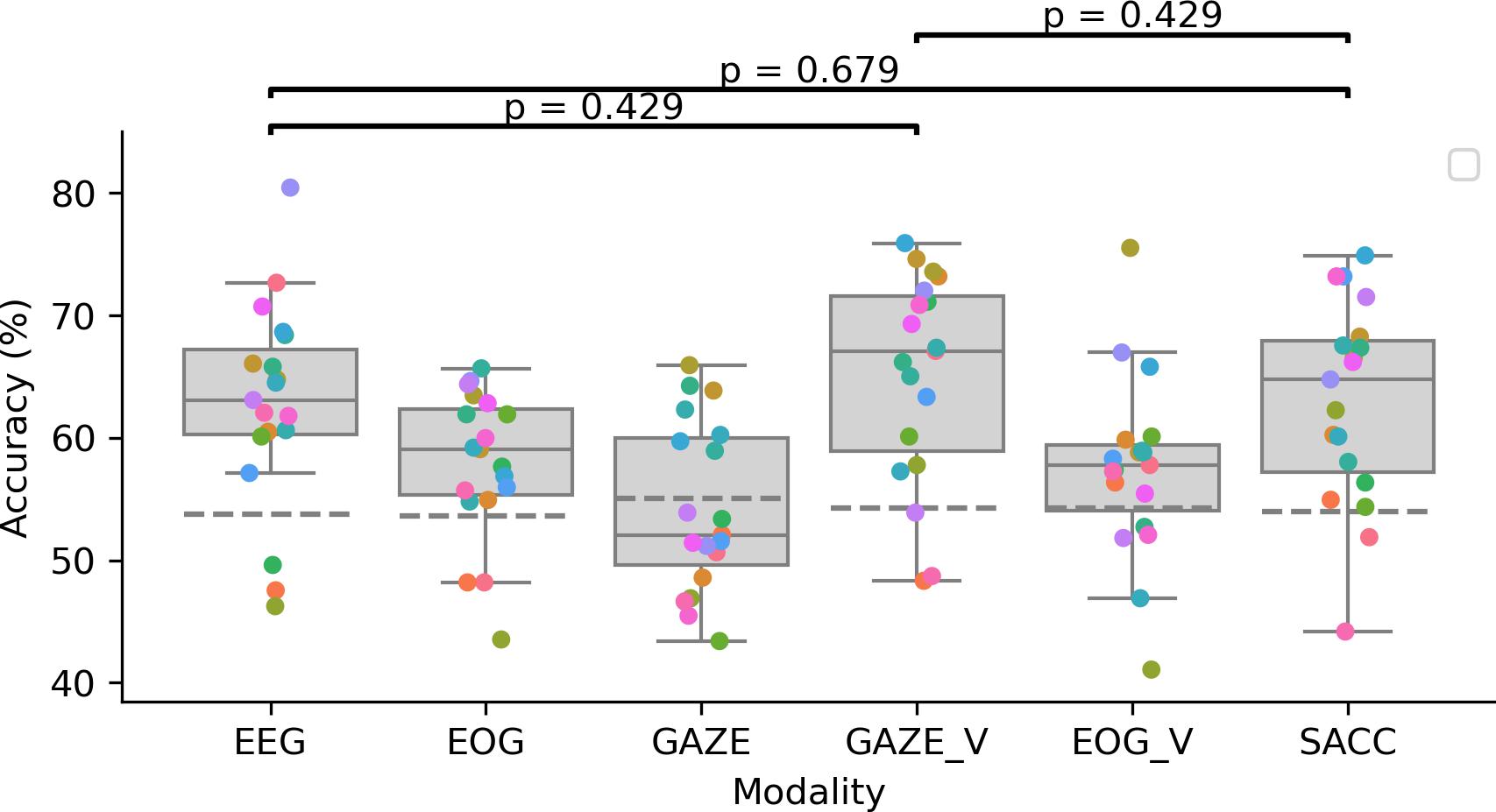}
  \caption{Accuracies of the SVAD task, i.e., identifying the attended video segment from the unattended video segment using data segments from different modalities. The models are trained and tested on the superimposed-object dataset and the test segments are \SI{30}{\second} long. The dots in different colors represent the accuracies of individual subjects. The boxes show the median and the interquartile range, with the whiskers extending to the most extreme data points. Wilcoxon signed-rank tests are performed to determine if the distributions are significantly different, and the p-values are indicated on the figure (BH-adjusted). The significance levels are indicated by the dashed lines.}
  \label{fig:VAD_res}
\end{figure}

  The results are shown in Fig. \ref{fig:VAD_res}. Among the tested modalities, $EEG$, $GAZE\_V$, and $SACC$ stand out with higher decoding accuracies, with approximately $75\%$ of subjects reaching $60\%$ accuracy or higher, and medians around $63.0\%$, $67.1\%$, and $64.8\%$, respectively. There is no significant difference in performance when comparing these three modalities.

\par Although the main focus of this study is EEG-based SVAD, it is noteworthy that this decoding can be achieved at least equally well using the gaze velocity or saccade information obtained from an eye tracker. The good performance of gaze velocity and saccade indicates that specific eye movement patterns elicited by the video stimuli can be informative for SVAD, even when the objects are spatially overlapping. The superior performance of gaze velocity amplitude over original gaze coordinates may be attributed to the fact that the \textit{ObjFlow} feature is also based on (pixel) velocity magnitude. 

\subsection{EEG-based decoding is not dominantly driven by eye movement artefacts} \label{sec:EEG_eye_mov}

\par In Fig. \ref{fig:VAD_res}, we notice that EEG does not outperform gaze velocity and saccades, which raises an important question: does EEG-based decoding primarily rely on eye movement artefacts in the EEG recordings? For some use cases, it may not be necessary to disentangle the effects of eye movements, as the primary goal is high decoding accuracy. However, since we also aim to enable novel experimental paradigms in neuroscience, where the focus is on neural responses, it is crucial to understand the role of eye movements in EEG-based decoding under free-viewing conditions. In this section, we suppress the effects of eye movement artefacts in the EEG (including saccades) in three ways: by regressing out eye movements (Section \ref{sec:eye_mov_regression}), by using only EEG channels in the visual cortex (Section \ref{sec:visual_cortex}), and by analyzing data free from saccades (Section \ref{sec:saccade}). The decoding accuracies are recomputed on the superimposed-object dataset under these three cases, collectively suggesting that EEG-based decoding is largely independent of eye movements.

\subsubsection{Visual attention is decodable after regressing out eye movements} \label{sec:eye_mov_regression}

\par
\begin{figure}
  \centering
  \includegraphics[width=0.7\linewidth]{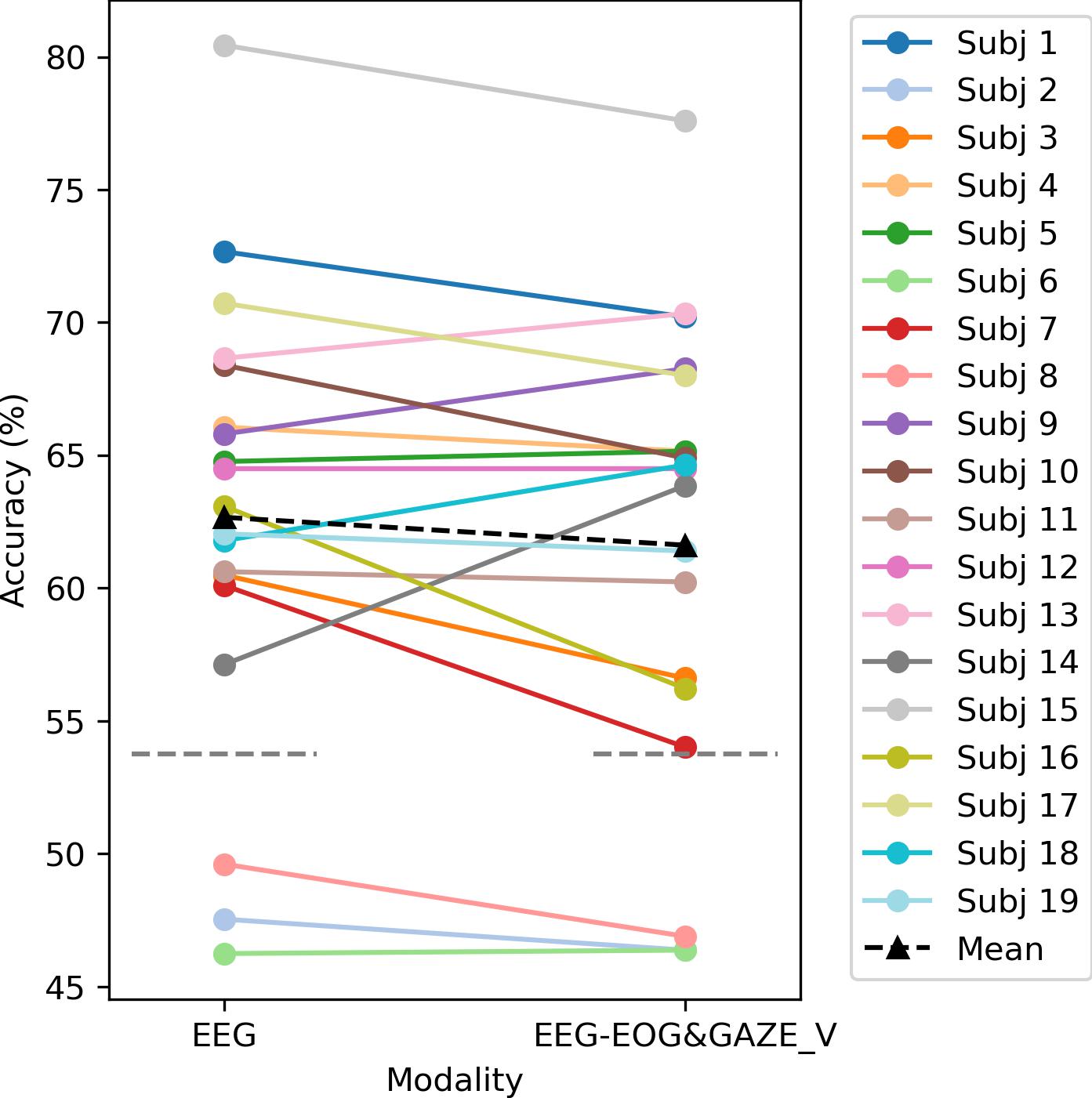}
  \caption{Accuracies of the SVAD task before and after regressing out EOG and gaze velocity from EEG. The latter is denoted by ``$EEG-EOG\&GAZE\_V$". The models are trained and tested on the superimposed-object dataset and the test segments are \SI{30}{\second} long. The dots denote the individual (per-subject) accuracies, and the results of the same subject are connected with a line. The significance levels are indicated by the dashed lines.}
  \label{fig:eye_mov_regression}
\end{figure}

 To control for the effects of eye movements, we apply PCCA (Section \ref{sec:pcca}) to correlate EEG signals with video features, setting the EOG and gaze velocity as confounds. More specifically, EOG and gaze velocity are concatenated along the channel axis and regressed out from both the EEG signals and the attended/unattended video features as in \eqref{eq:residuals}. CCA is then applied to the residuals to find the canonical components. Note that saccade information, which is encoded in gaze velocity as sudden changes in coordinates leading to peaks in velocity (Section \ref{sec:data_preprocessing}), is also implicitly suppressed after regression. The accuracies before and after controlling for eye movements are shown in Fig. \ref{fig:eye_mov_regression}. Although decoding performance declines significantly after regression (p-value = $0.049$), most subjects exhibit modest changes, with average accuracy decreasing slightly from $62.7\%$ to $61.6\%$. Furthermore, in subjects with significant decoding accuracy, this significance persists even after regressing out gaze information. This result suggests that while eye movement information in EEG can assist SVAD, it does not primarily drive the decoding performance.

\par However, a limitation of the above analysis is that the ``eye movement information" considered here includes only EOG and gaze velocity, and not all information related to eye movements. For instance, there could be other nonlinear transformations of the gaze coordinates that correlate with visual stimuli and EEG signals, thereby boosting the decoding performance. Another way to disentangle or reduce the influence of eye movements is to use EEG channels that are less affected by eye movements, which is discussed in the next section.

\subsubsection{Visual attention is decodable using channels in visual cortex} \label{sec:visual_cortex}

\par
\begin{figure*}
  \centering
  \begin{subfigure}[b]{0.35\textwidth}
    \centering
    \includegraphics[width=\linewidth]{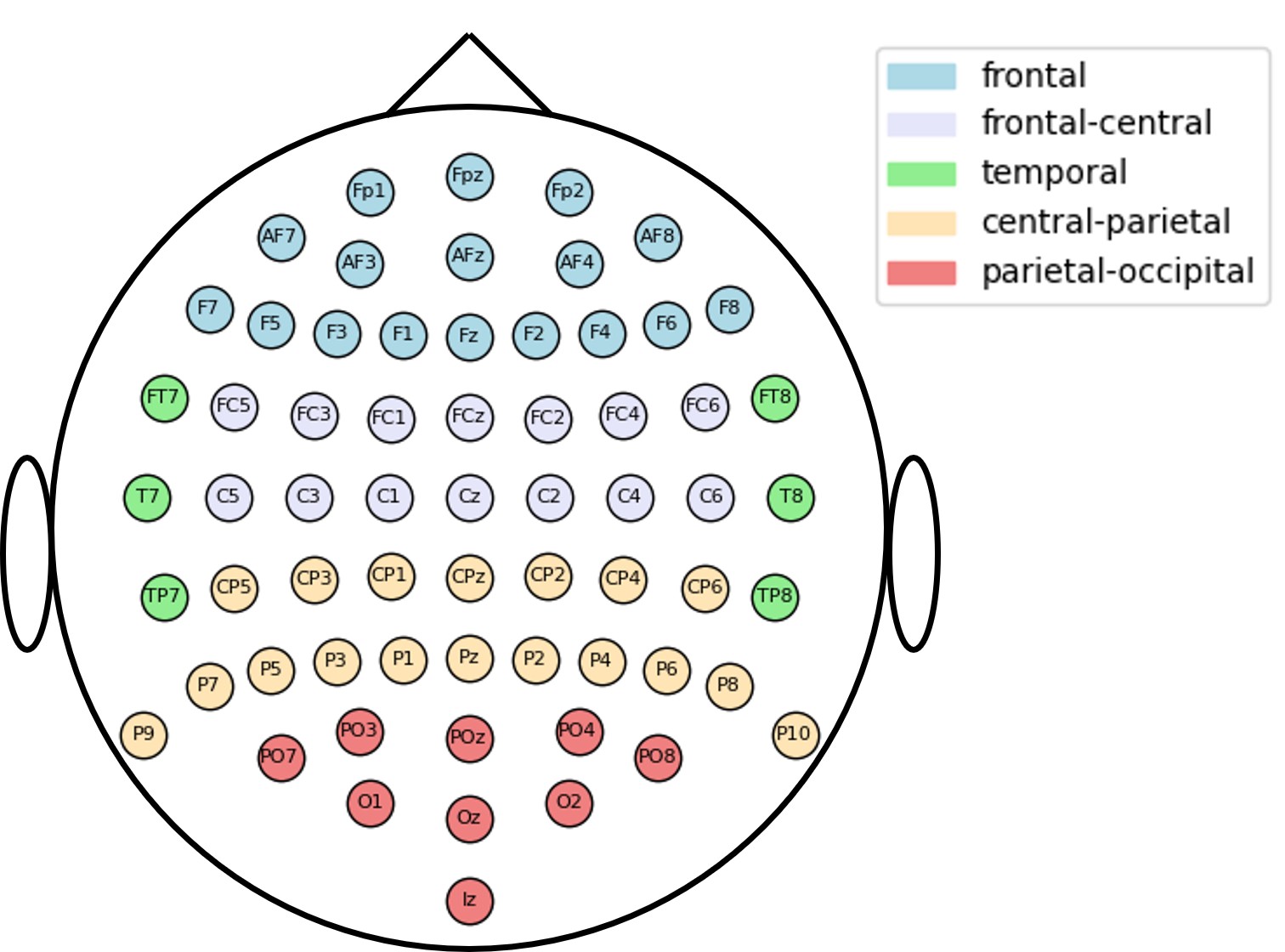}
    \caption{}
    \label{fig:regions_EEG}
  \end{subfigure}
  \hfill
  \begin{subfigure}[b]{0.62\textwidth}
    \centering
    \includegraphics[width=\linewidth]{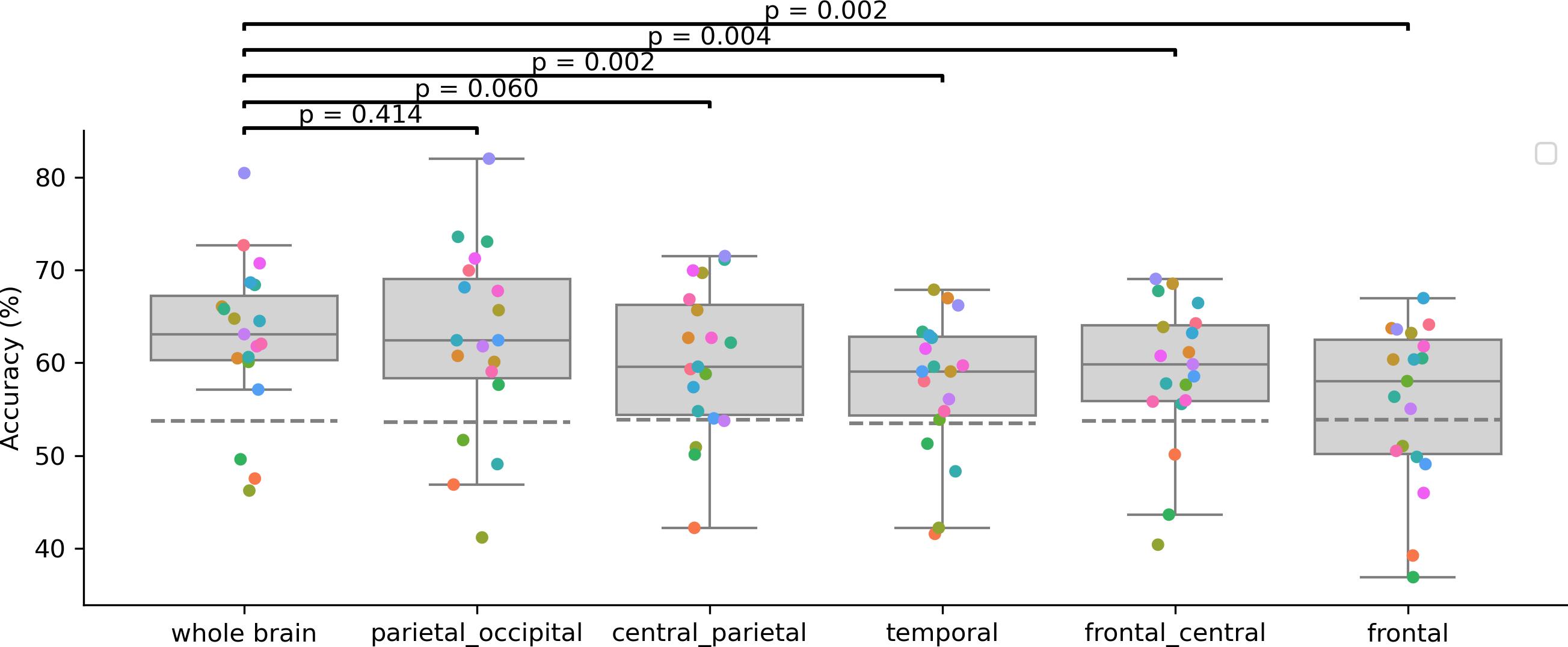}
    \caption{}
    \label{fig:regions_res}
  \end{subfigure}
  \caption{  (a) Topographic maps of the EEG channels in different brain regions. (b) Accuracies of the SVAD task using EEG channels in different brain regions. The models are trained and tested on the superimposed-object dataset and the test segments are \SI{30}{\second} long. The dots denote the individual (per-subject) accuracies, and the boxes show the median and the interquartile range. Wilcoxon signed-rank tests are performed to determine if using whole-brain signals is significantly better than using signals from a specific region. The p-values are BH-adjusted. The significance levels are indicated by the dashed lines.}
  \label{fig:combined_regions}
\end{figure*}

 It is known that eye movements primarily affect EEG channels in the frontal region, with the strength of these artefacts decreasing as they propagate towards the back of the head \cite{romero2008comparative}. Therefore, if EEG-based decoding is mainly driven by eye movements, the decoding performance when using channels in the frontal region should be better compared to other regions, especially the parietal-occipital region, where the visual cortex is located, being the furthest away from the eyes. To test this hypothesis, we divide EEG channels into different groups based on their locations (Fig. \ref{fig:regions_EEG}), and perform the SVAD task using each group. The results are presented in Fig. \ref{fig:regions_res}. Contrary to the hypothesis, the decoding accuracy using channels in the parietal-occipital region is comparable to the performance using whole-brain signals, whereas accuracy gradually declines in regions closer to the eyes. This result suggests that EEG-based decoding is primarily driven by neural responses in the visual cortex rather than eye movements.

\subsubsection{Visual attention is decodable after removing saccades} \label{sec:saccade}

\par Saccades also elicit neural responses, which have been found to be dominant across the entire brain under a free-viewing setup similar to ours \cite{nentwich2023semantic}. As mentioned in Section \ref{sec:eye_mov_regression}, the effect of saccades is suppressed after regressing out EOG and gaze velocity from EEG and video features. However, a safer option is to remove the data segments around saccades and analyze the remaining data, as regression might not fully eliminate the event-related potentials elicited by saccades.

\par In our experiment, we remove data points from \SI{0.33}{\second} before to \SI{1}{\second} after the saccade onset, resulting in a data loss ranging from $23\%$ to $82\%$, depending on the subject. To ensure sufficient data for training and testing, we train the CCA decoders on the single-object dataset in a subject-independent manner (i.e. concatenating the data from all subjects) and test on the superimposed-object dataset for individual subjects. For a fair comparison, we create control groups by randomly removing the same amount of data not necessarily around saccades. Wilcoxon signed-rank tests are performed to determine if the performance after removing data around saccades is significantly worse than the control groups. The BH-adjusted p-values are all above $0.05$ ($0.276$, $0.410$, $0.252$, $0.225$, $0.225$, $0.252$, $0.414$, $0.225$, $0.225$, $0.225$), indicating no strong evidence that EEG-based decoding is primarily driven by saccades.


\subsection{EEG may also capture information of the unattended object} \label{sec:VAD_MM}

\par In Section \ref{sec:correlation_modulation}, we observe that the correlations with the unattended object are not only lower but also non-significant for all data modalities. This is remarkable, especially for EEG, since the unattended object is present in the same location in the visual field as the attended object. This suggests that the brain is able to separate the visual streams of both objects and suppress one of them in favor of the other. Since the correlation with the unattended object is not significant, the question remains whether the EEG actually contains signal components that encode the unattended object and whether these can be captured by the CCA model. To address this question, we conduct the match-mismatch (MM) task (Section \ref{sec:evaluation_tasks}), where the attended video segment remains the same as in SVAD, and the unattended video segment is an unobserved segment at a random time point in the same test set. We apply the same bootstrapping and cross-validation procedure, with video segment lengths still set to \SI{30}{\second}, and compare the decoding accuracies of the SVAD and MM tasks on the superimposed-object dataset.

\par
\begin{figure*}
  \centering
  \begin{subfigure}[b]{0.43\textwidth}
    \centering
    \includegraphics[width=\linewidth]{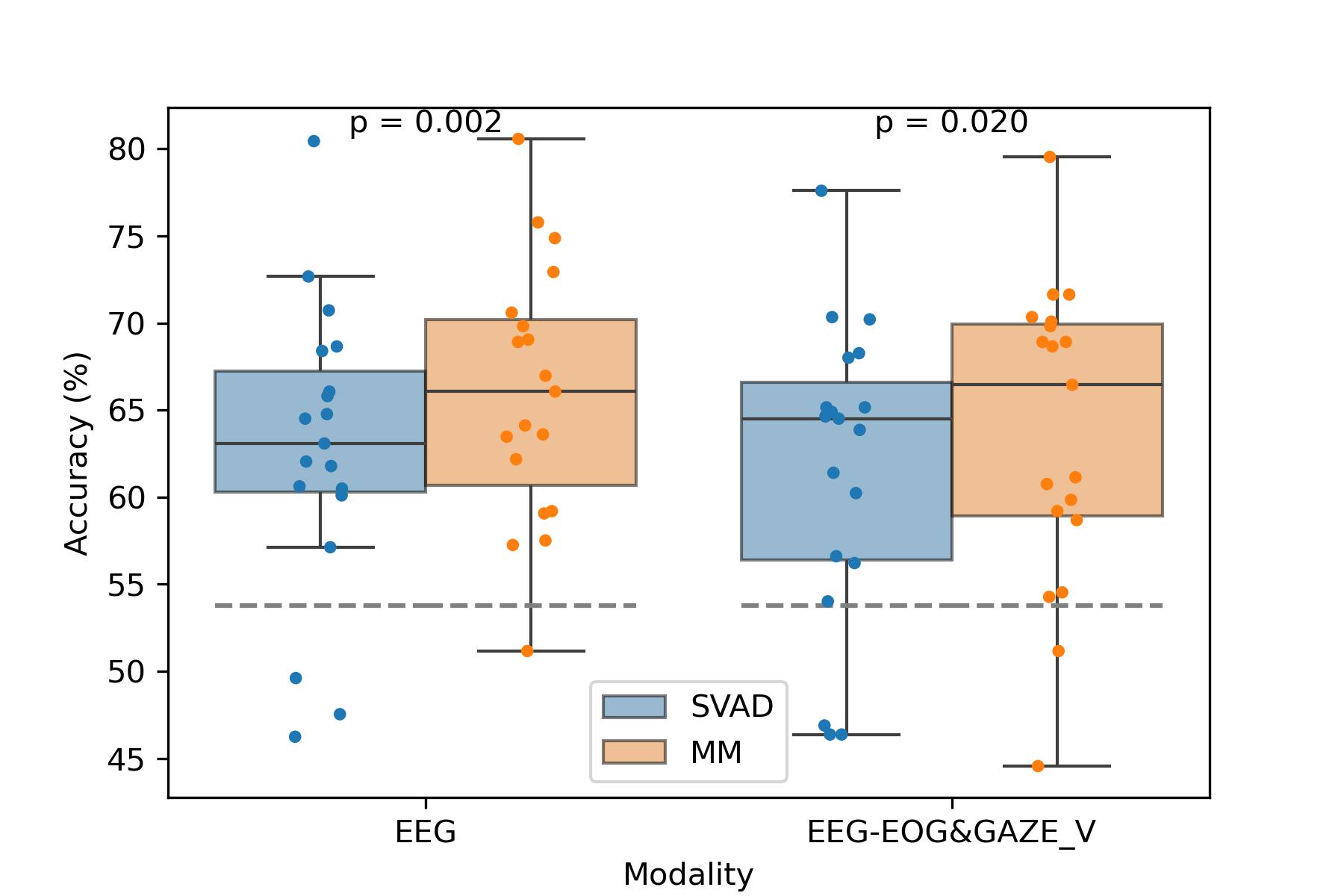}
    \caption{}
    \label{fig:VAD_MM_acc}
  \end{subfigure}
  \hfill
  \begin{subfigure}[b]{0.5\textwidth}
    \centering
    \includegraphics[width=\linewidth]{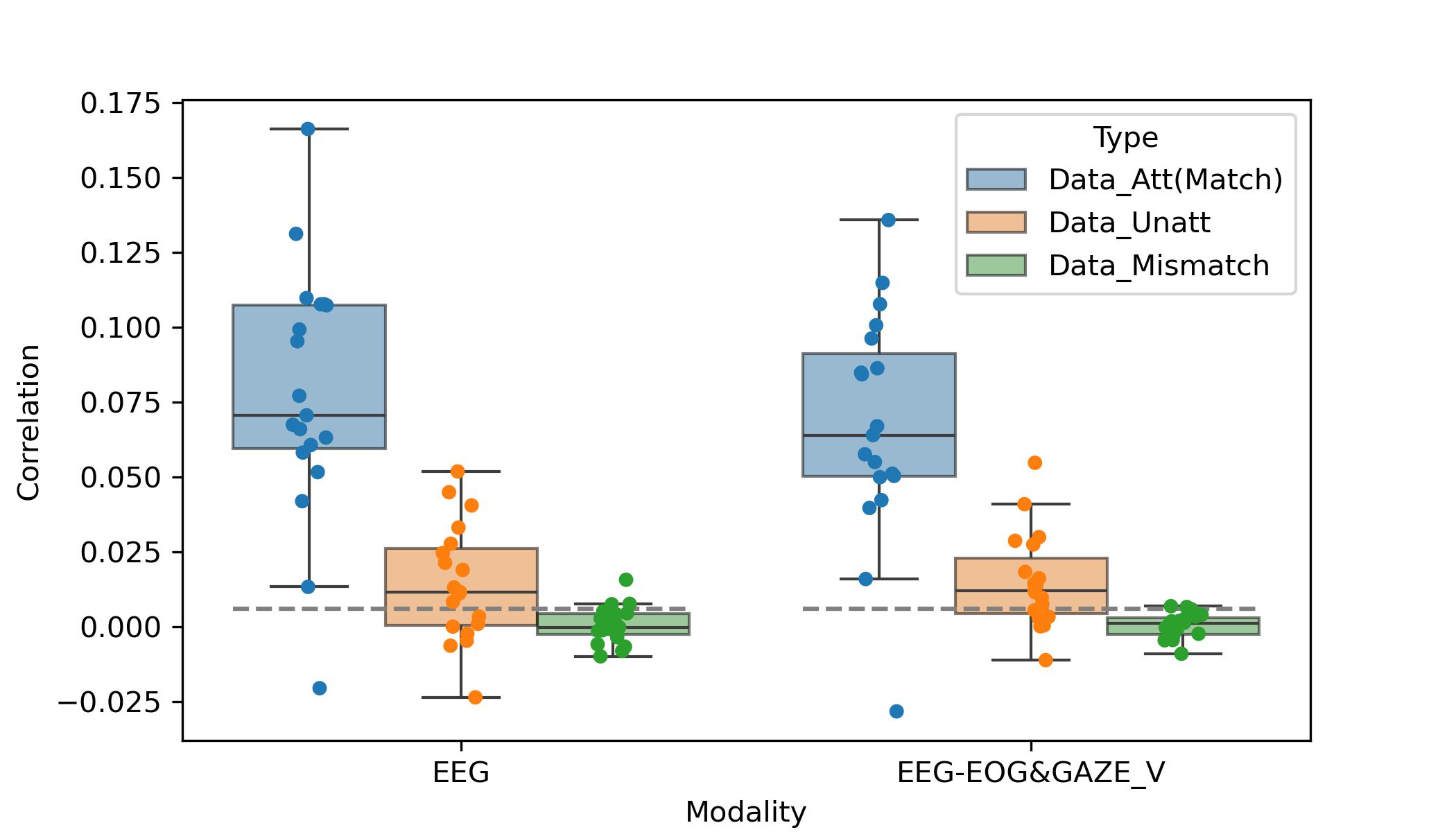}
    \caption{}
    \label{fig:eeg_att_unatt_mm}
  \end{subfigure}
  \caption{  (a) Accuracies of the SVAD and MM tasks. Wilcoxon signed-rank tests are performed to determine if the accuracies of SVAD tasks are significantly lower than those of the MM tasks using EEG. The p-values are BH-adjusted. The significance levels are indicated by the dashed lines. (b) The sum of first two canonical correlations between each modality and the \textit{ObjFlow} feature of the attended object (in blue), the unattended object (in orange) and the mismatch object (in green). The models are trained and tested on the superimposed-object dataset and the test segments are \SI{30}{\second} long. The dots denote the per-subject results averaged across all trials, and the boxes show the median and the interquartile range. The significance levels are indicated by the dashed lines.}
  \label{fig:combined_vad_mm}
\end{figure*}

 The results are shown in Fig. \ref{fig:VAD_MM_acc}. Both before and after regressing out eye movements, the EEG-based decoding accuracies of the MM task are significantly higher than those of the SVAD task, despite the overall difference being small. This indicates that EEG may capture information about the unattended video, which confuses the SVAD decision. Further evidence is provided by the results in Fig. \ref{fig:eeg_att_unatt_mm}, which show the sum of the first two canonical correlations between EEG (with or without eye movement regressed out) and the \textit{ObjFlow} feature of the attended, unattended, and mismatch object. The significant difference between the correlations with the unattended object versus the mismatch object implies that the model also extracts responses correlated with the unattended object. Note that for eye-related data modalities, the performance of the MM task is comparable to or even worse than that of the SVAD task, suggesting that the unattended object is hardly captured by these modalities (Supplementary Material \cite{yao_2025_15211457}, Section I).


\subsection{Complementary information exists in EEG and gaze features}\label{sec:complementary_info}

\par
\begin{figure}
  \centering
  \includegraphics[width=.95\linewidth]{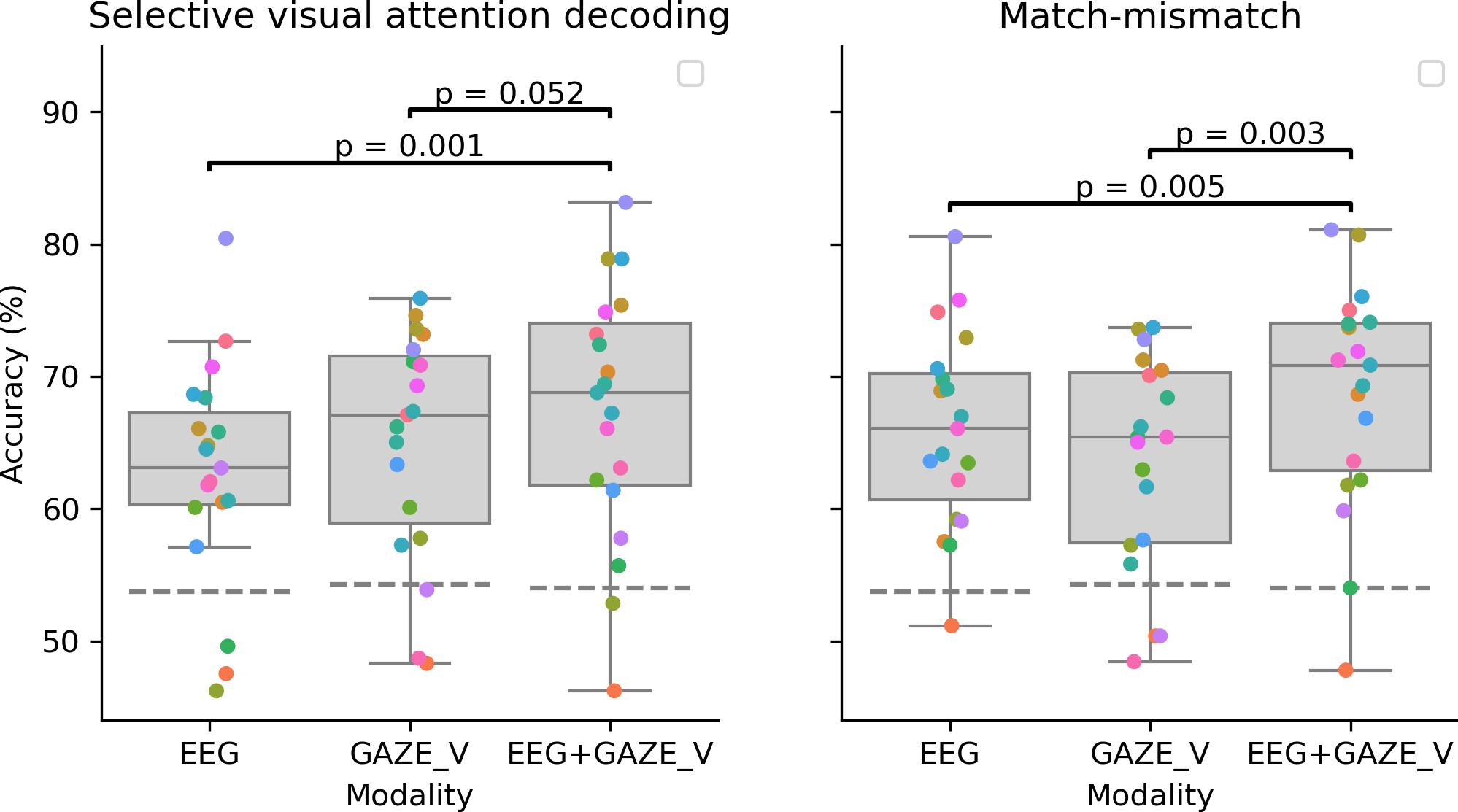}
  \caption{  Accuracies of the SVAD and MM tasks using EEG, gaze velocity, and their combination. The models are trained and tested on the superimposed-object dataset and the test segments are \SI{30}{\second} long. The dots denote the individual (per-subject) accuracies, and the boxes show the median and the interquartile range. Wilcoxon signed-rank tests are performed to determine if the accuracies of using EEG and gaze velocity separately are significantly lower than that of using them together. The p-values are BH-adjusted. The significance levels are indicated by the dashed lines.}
  \label{fig:complementary_info}
\end{figure}

 In Section \ref{sec:eye_mov_regression}, we have demonstrated that regressing out gaze velocity (and EOG) from EEG signals does not drastically affect decoding performance. Therefore, it is reasonable to assume that these two data modalities capture complementary information. This raises a natural question: can combining them improve decoding performance? The combination can be achieved by simply concatenating gaze velocity to EEG signals as an extra channel. We then apply the decoding pipeline to the combined data and compare the performance with using EEG and gaze velocity separately. The accuracies of the two tasks on the superimposed-object dataset using EEG, gaze velocity, and their combination are shown in Fig. \ref{fig:complementary_info}. 

\par In the SVAD task, using combined modalities significantly outperforms using EEG alone but not gaze velocity alone, with a median accuracy around $68.8\%$. In the MM task, the performance of using combined modalities is significantly higher than using each of them separately, with a median accuracy around $70.9\%$. These results suggest that the information captured by EEG and gaze velocity is complementary and can lead to better decoding performance, especially in the MM task. However, for the SVAD task, the additional information from EEG may not be as discriminative as the information in gaze velocity (as already explained in Section \ref{sec:VAD_MM}), leading to limited improvement.

\subsection{Synchronization among subjects decreases in the presence of a distractor} \label{sec:GCCA_res}

\par In the previous sections, we have focused on stimulus-aware individual-level analysis, correlating video features with data modalities and identifying the attended video segment. Now, we shift our focus to group-level analysis, which bypasses video feature extraction, quantifies the synchronization level among participants, and provides insights into group attention or engagement \cite{dmochowski2012correlated, madsen2021synchronized, geirnaert2023stimulusInformed}. Specifically, we apply GCCA (Section \ref{sec:gcca}) to each data modality and compute the inter-subject correlation (ISC) for the first canonical component for both datasets: single-object and superimposed-object. The ISCs are cross-validated using a leave-one-pair-out scheme. 

\par
\begin{figure*}
  \centering
  \includegraphics[width=0.7\linewidth]{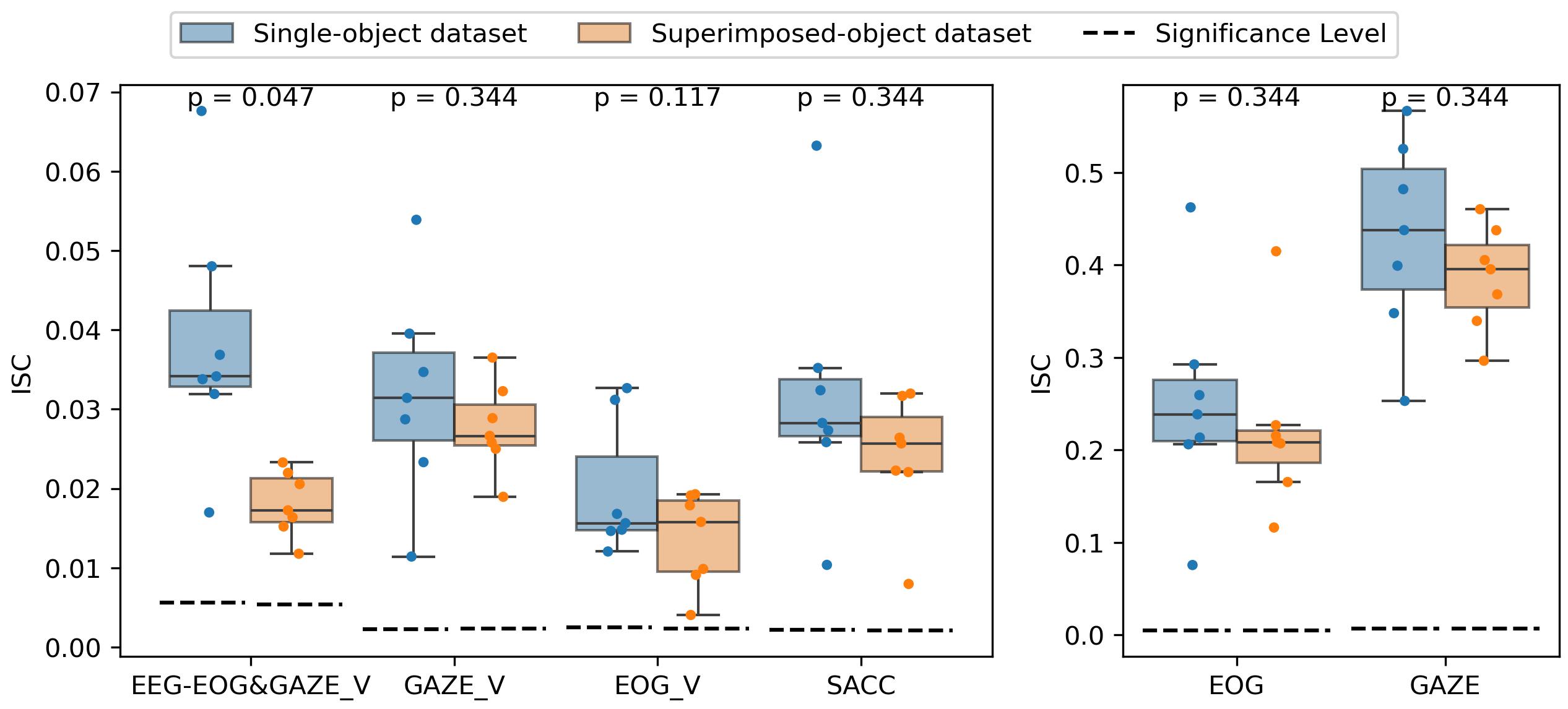}
  \caption{  Inter-subject correlations (ISCs) of the first canonical component across different data modalities for the single-object dataset (blue) and the superimposed-object dataset (orange). The ISCs for $EOG$ and $GAZE$ are plotted separately because their correlation values are much higher due to their particular signal characteristics, i.e., they are much lower in frequency and are approximately piecewise constant. Each dot represents the ISC for an individual fold, and the boxes display the median and interquartile range. Wilcoxon signed-rank tests are performed to assess whether ISCs in the single-object dataset are significantly higher than those in the superimposed-object dataset, with p-values adjusted using the BH procedure.}
  \label{fig:ISC_res}
\end{figure*}

 The results for each fold are presented in Fig. \ref{fig:ISC_res} \footnote{Note that ISCs should not be compared across modalities as they heavily depend on the spectral characteristics and signal-to-noise ratio of the underlying signals.}, from which we can observe that the ISCs of EEG (with eye movements regressed out) are significantly lower in the superimposed-object dataset. A decreasing trend in ISCs is also evident for the other modalities, although this decrease is not statistically significant, potentially due to the limited sample size. This indicates that synchronization among subjects decreases when a distractor is present. A possible explanation is that subjects may be distracted by the unattended object and this distraction can happen at different points in time for different subjects, leading to a reduced synchronicity. This poses a challenge for stimulus-unaware ISC-based measurements of attention when viewing naturalistic videos: the attention of the participants is more scattered in the presence of multiple objects, and lower ISCs do not necessarily imply lower absolute attention levels to the overall stimuli. For example, in a tennis match recording, participants might focus on different players at different times, resulting in lower ISCs even if their attention levels are high. Another interesting observation is that despite the high synchronization of EOG and gaze coordinates across subjects, they do not perform well in the SVAD and MM tasks (Fig. \ref{fig:VAD_MM_acc}), whose performance depends more on the correlation between the data and the extracted video features.

\section{Discussion} \label{sec:discussion}

\subsection{Is the ground truth reliable?}

\par In this study, we have assumed that participants always follow the instruction, and we use the object they are asked to attend to as the ground truth in the SVAD task. Although this may not always be the case, we expect the ground truth to remain reliable assuming participants only occasionally attend to the distractor.

\par
\begin{figure}
  \centering
  \includegraphics[width=\linewidth]{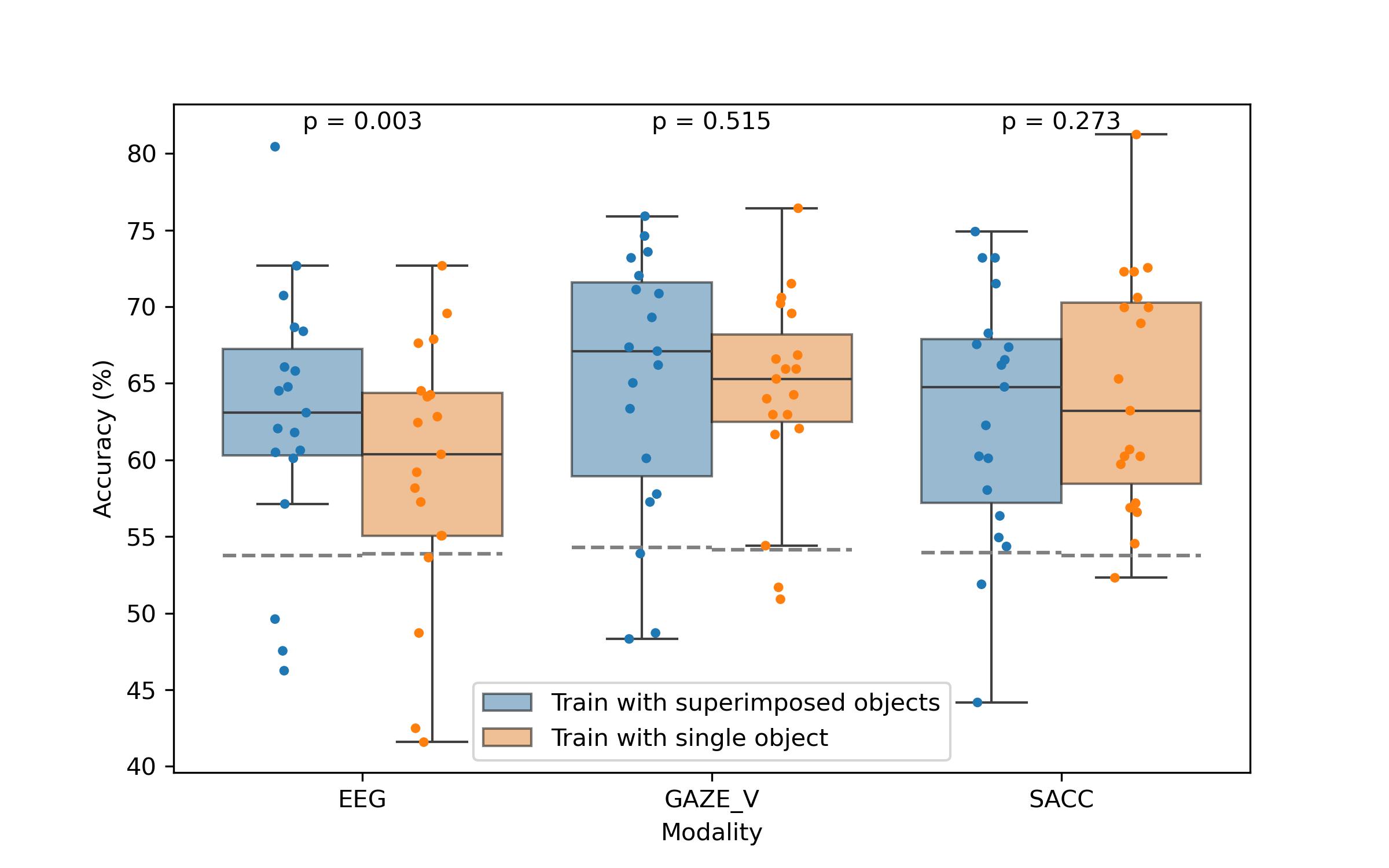}
  \caption{ Accuracies of the SVAD task using decoders trained in the superimposed-object dataset (in blue) and the single-object dataset (in orange). The test segments are \SI{30}{\second} long. The dots denote the individual (per-subject) accuracies, and the boxes show the median and the interquartile range. Two-sided Wilcoxon signed-rank tests are performed and the p-values are BH-adjusted. The significance levels are indicated by the dashed lines.}
  \label{fig:acc_vad_SO}
\end{figure}

 Additional evidence for this assumption can be found in Fig. \ref{fig:acc_vad_SO}. Here, we repeat the analysis described in Section \ref{sec:VAD_res} for the best three modalities where the CCA decoders are this time trained on the single-object dataset, for which the ground truth is certain due to the absence of a distracting object. From Fig. \ref{fig:acc_vad_SO}, we conclude that the impact is relatively mild; for the $GAZE\_V$ and $SACC$ modalities the difference is not significant, and in the case of $EEG$, training with the single-object data actually leads to significantly worse results, despite the availability of an exact ground truth. In this case, training with superimposed objects results in higher accuracies, which would be unlikely if the ground truth in this data set would be unreliable.

\par The decrease in performance when training the decoders on the single-object data in the case of EEG might be explained by the fact that the decoder can not learn to suppress neural responses to the unattended object, as these responses are not present in the training set. Another possible explanation is the fact that the task of attending to a target object in the superimposed videos is more challenging, which could result in stronger neural responses, which are more easily decodable. A similar effect has been described in the context of selective attention decoding with speech stimuli, where more difficult tasks, i.e., in more challenging acoustic conditions, can result in better decoding accuracies \cite{das2016effect, das2018eeg}.

\subsection{Eye tracker or EEG}

\par In selective visual attention decoding, eye trackers are perhaps a more straightforward and popular choice. Their advantage in overt attention decoding is evident as they directly provide gaze information with high spatial and temporal resolution. Therefore, gaze maps exported from eye trackers are usually considered the ground truth for attention in many fields such as video saliency prediction \cite{jiang2018deepvs}, neuromarketing \cite{dos2015eye}, and cognitive workload measurement \cite{zagermann2016measuring}. Additionally, eye trackers have been found useful in quantifying the absolute attention level. For example, in \cite{madsen2021synchronized}, ISCs of gaze and pupil size were used as markers of attention and were predictive of students' test scores on a group-level.

\par A limitation of eye trackers is that they cannot measure covert attention, which can be problematic when participants attend to objects in their peripheral vision without moving their eyes. However, this is less of a concern in free-viewing scenarios where overt attention is more prevalent. Eye trackers may also struggle when objects are close to each other, as the gaze coordinates might be ambiguous. Nevertheless, in this study we demonstrated that gaze velocity and saccades can still be informative even when objects are spatially overlapped, as long as the objects have distinct motion patterns.

\par Another limitation of eye trackers in the context of (selective) attention decoding is that eye movements are an indirect measure of attention since there is a gap between merely ``looking at" and ``paying attention", whereas EEG directly captures neural responses to stimuli, allowing for arguably more reliable inference of attention, independent of gaze patterns. This could be particularly useful when one aims to decode not only the attended object but also how attentive the participant is to the object. Take the results of Subject 11 and 17 (Table \ref{tab:subj_acc_comparison}) as an example: the gaze velocity-based decoding accuracy is comparable for both subjects, but the EEG-based decoding accuracy (with eye movements regressed out) is $8\%$ lower for Subject 11. Additionally, the EEG signals of Subject 11 exhibit higher correlations with the unattended object. This suggests that Subject 11 may track the attended object similarly to Subject 17 but is less successful in suppressing the distractor that spatially overlaps with the target object. Together with the fact that the correlation between EEG and the attended object is also lower for Subject 11, this indicates that Subject 11 might be less attentive overall than Subject 17.

\begin{table}[h!]
  \centering
  \caption{  Comparison of accuracies between gaze velocity-based and EEG-based decoding (with eye movements regressed out) in the SVAD task on the superimposed-object dataset for two selected subjects. The sum of the first two canonical correlations between EEG and attended video features, and EEG and unattended video features are also shown.}
  \label{tab:subj_acc_comparison}
  \begin{tabular}{cccccc}
  \toprule
  \textbf{Subject} & \multicolumn{2}{c}{\textbf{Accuracy (SVAD, \%)}} & \multicolumn{2}{c}{\textbf{Correlation}} \\
  \textbf{ID} & \textbf{Gaze Velocity-} & \textbf{EEG-Based} & \textbf{EEG-Att} & \textbf{EEG-Unatt} \\
  & \textbf{Based} &  & ~ & ~\\
  \midrule
  11 & 67.4 & 60.2 & 0.058 & 0.041 \\
  17 & 69.3& 68.0 & 0.097 & 0.011 \\
  \bottomrule
  \end{tabular}
\end{table}

\par Neural-based decoding can also be advantageous for understanding the underlying neural mechanisms of selective attention, such as the timing of attentional effects and sources of attentional control signals \cite{kellerAEC2022, grootswagersND2021, goddard2022spatial}, although EEG-based paradigms are not yet prevalent. While current decoding accuracies may not yet meet the requirements for real-world applications, they can be improved by reducing the time resolution (i.e. use longer test windows) or incorporating evidence-accumulation techniques such as hidden Markov models or state-space models, as sometimes used in speech decoding \cite{geirnaert2019interpretable, aroudi2020improving, heintz2024probabilistic}.

\subsection{Gaze-informed attention decoding}

\par In Section \ref{sec:complementary_info}, we have combined EEG and gaze velocity by concatenating them along the channel axis. Another way to incorporate gaze information is during video feature extraction, by weighting or selecting video features based on gaze coordinates. The goal is then not only to identify the attended object but also to measure the absolute attention level to the object in the gaze direction. This approach is motivated by the selective attention mechanism: we assume participants direct their gaze to the object of interest in free-viewing scenarios, and features around the gaze coordinates should be emphasized to find strong correlations with EEG signals, as attended features are enhanced in the brain. This is particularly useful when multiple objects are interacting in the scene and each participant's attention is unknown beforehand. A region defined based on the gaze map could potentially replace the bounding boxes of objects in the object-based features proposed in \cite{yao2024identifying}, circumventing the need for feature fusion in multi-object scenarios. 

\subsection{Gaze-driven EEG components versus stimulus-driven neural responses}

\par Returning to the proposed EEG-based decoder, although the analysis in Section \ref{sec:EEG_eye_mov} shows that the decoding performance is unlikely to be driven by eye movements, we cannot make definitive arguments as not all confounds are completely removed. Methods of eye movement artifact removal, such as the linear regression in our PCCA procedure, only suppress the artefacts without necessarily fully eliminating them, and the residuals might still affect decoding performance. Analyzing only EEG channels in the occipital-parietal region also mitigates ocular contamination, but neural responses elicited by saccades are still present in that region \cite{nentwich2023semantic}. Cutting out segments around saccades appears to be a reliable way of obtaining ``clean" data, but the resulting discontinuities in the signals might introduce new confounds. Additionally, it is difficult to disentangle the EEG signals related to the motor control of eye movements.

\par In essence, eye movement is a complex process that involves multiple brain regions and is closely linked to attention. Consequently, it is not feasible to fully disentangle all related confounds. While the results of this study suggest that EEG-based decoding is not dominantly driven by eye movements, we acknowledge that, depending on the research question and application area, gaze fixation may be necessary to better disentangle the effects of eye movements from EEG signals, as opposed to the free-viewing paradigm that was used in this study.

\section{Conclusion} \label{sec:conclusion}

\par In this study, we propose an experimental protocol for selective visual attention decoding that better approximates real-world conditions—though not fully replicating them—by introducing naturalistic videos over synthetic or static images, allowing free-viewing rather than enforcing gaze fixation, and utilizing EEG instead of fMRI. We demonstrate that it is possible to decode the attended object from EEG signals, even when using only visual cortex channels and when the two objects are co-located, thereby ruling out position-based confounds. We provided supporting empirical evidence that the neural tracking of naturalistic motion is modulated by selective attention. Apart from EEG, eye gaze data have also been used to decode attention. Although the attended and unattended objects are superimposed, the target is still decodable from gaze data since the gaze velocity and saccades are related to the movement pattern of the attended object. 

\par To better understand the role of eye movements in EEG-based decoding, we have conducted a series of experiments to disentangle possible eye gaze confounds in the EEG signals. The results indicate that EEG-based decoding is not dominantly driven by eye movements. We have also demonstrated that EEG likely captures information about both the attended and unattended objects, which makes the EEG-based decoder less discriminative. This finding may explain why adding EEG to gaze data does not significantly improve SVAD performance, despite the existence of complementary information. Furthermore, group-level analysis reveals that the participants' attention is more scattered when a distractor is present, making stimulus-unaware group-level attention metrics such as ISC less reliable with increased stimulus complexity.

\par As a first study of selective visual attention decoding in natural videos using EEG, it takes the middle ground between experimental control and ecological validity. Future work could proceed in two directions. First, a more controlled approach could replicate this experiment with fixed gaze protocols to fully isolate ocular activities. Alternatively, a more application-focused direction could go for more ecological setups, identifying more relevant video features, developing more sophisticated models, or integrating gaze-informed decoding strategies to improve performance.

\section*{References}

\bibliographystyle{IEEEtran}  
\bibliography{references} 

\end{document}